\documentclass[%
 reprint,
 amsmath,amssymb,
 aps, 
 pra,
 showkeys, 
 superscriptaddress
]{revtex4-2}
\usepackage{graphicx}
\usepackage{dcolumn}

\usepackage[main=english]{babel}
\usepackage[utf8]{inputenc}
\usepackage{csquotes}

\usepackage{float}
\usepackage{enumitem}
\usepackage{comment}
\usepackage{bm}

\usepackage{graphicx}
\usepackage[export]{adjustbox}

\usepackage{amsmath}
\usepackage{amsfonts}
\usepackage{amssymb}
\usepackage{mathrsfs}
\usepackage{mathtools}
\usepackage{stmaryrd}

\usepackage{amsthm}

\theoremstyle{definition}

\newcommand{\diff}{\mathop{}\!\mathrm{d}}
\newcommand{\difft}{\frac{\diff }{\diff t}}

\newcommand{\daggah}{^\dagger}
\usepackage{dsfont}

\newcommand{\timeorder}{\mathcal{T}_+}

\DeclarePairedDelimiter{\bra}{\langle}{|}
\DeclarePairedDelimiter{\ket}{|}{\rangle}
\newcommand{\ketbra}[2]{\ket{#1}\bra{#2} }
\newcommand{\braket}[2]{\langle{#1}|{#2}\rangle }

\renewcommand{\Im}{\mathrm{Im}}

\usepackage{xspace}
\newcommand{\schr}{Schr\"{o}dinger\xspace}
\newcommand{\ito}{It\={o}\xspace}

\usepackage[dvipsnames]{xcolor}
\newcommand{\deepgreen}[1]{\textcolor{teal}{#1}} 
\newcommand{\deepblue}[1]{\textcolor{Blue}{#1}}

\usepackage{tikz}
\usetikzlibrary{arrows.meta,bending,chains}
\usetikzlibrary{calc}
\usetikzlibrary{angles} 
\usetikzlibrary{positioning}
\usetikzlibrary{hobby}
\usetikzlibrary{backgrounds,fit,decorations.pathreplacing}  
\usetikzlibrary{quantikz2}

\DeclareUnicodeCharacter{1D436}{}

\usepackage{blindtext}
\usepackage{hyperref}
\usepackage{cleveref}

\preprint{APS/123-QED}

\begin{document}

\title{On the Noisy Road to Open Quantum Dynamics: The Place of Stochastic Hamiltonians}
    
    \author{Pietro De Checchi}
    \email[Corresponding at: ]{pietro.dechecchi@math.unipd.it}
    \affiliation{Department of Mathematics, University of Padova, Via Trieste 63, Padova 35131,  Italy}%
    
    \author{Federico Gallina}
    \altaffiliation[Current affiliation: ]{Department of Physics, University of Ottawa, Ottawa, Ontario K1N 6N5, Canada}
    \affiliation{
     Department of Chemical Sciences, University of Padova, Via Marzolo 1, Padova 35131,  Italy}
    
    \author{Barbara Fresch}
    \affiliation{
     Department of Chemical Sciences, University of Padova, Via Marzolo 1, Padova 35131,  Italy}
    \affiliation{Padua Quantum Technologies Research Center, University of Padova, via Gradenigo 6/A, Padua 35131, Italy}
    
    \author{Giulio G.\ Giusteri}%
    \email[Corresponding at: ]{giulio.giusteri@unipd.it}
    \affiliation{Department of Mathematics, University of Padova, Via Trieste 63, Padova 35131,  Italy}%
    \affiliation{Padua Quantum Technologies Research Center, University of Padova, via Gradenigo 6/A, Padua 35131, Italy}
    
    \date{\today}

\begin{abstract}
    Stochastic evolution underpins several approaches to the dynamics of open quantum systems, such as random modulation of Hamiltonian parameters, the stochastic \schr equation (SSE), and the stochastic Liouville equation (SLE). 
    These approaches replace the explicit system–environment coupling with an effective system-only dynamics, where dissipative behavior emerges from ensemble averaging.
    Stochastic Hamiltonians, in particular, have long served as phenomenological tools in physical chemistry to include environmental effects without recourse to an explicit microscopic derivation.
    In this work, we aim at a self-contained and accessible presentation of these approaches to further elaborate on their common roots in essential concepts of stochastic calculus and to delineate the conditions under which they are equivalent. We also discuss how different formulations naturally lead to different numerical time-integration schemes, better suited for either classical simulation platforms, based on finite-difference approximations, or quantum algorithms, that employ random unitary maps. Our analysis aims at providing a unified perspective and actionable recipes for classical and quantum implementations of stochastic evolution in the simulation of open quantum systems.
\end{abstract}


\maketitle

\section{Introduction}\label{sec:intro}

Quantum systems are never fully isolated. Molecules and nano-structures, both natural and artificial, are embedded in some medium, such as solid-state matrices or fluids. Even almost-isolated systems in high vacuum, such as the neutral atoms and trapped ions used in quantum computing architectures, interact with the control apparatus and external fields. The surroundings and all the secondary degrees of freedom are part of the so-called environment of the system. The unavoidable interactions with the environment influence the expected ideal behavior of quantum systems, inducing decoherence and dissipation \cite{petruccione2002,Vacchini2024OpenTheory}.
An effective treatment of open quantum systems then requires shifting from the unitary evolution of a wavefunction to a density matrix description, whose evolution does not follow a universal equation of motion but instead depends on the approximations and modeling strategies that one employs. 

Two complementary perspectives are commonly adopted: in the first one, the density matrix describing the system of interest results from the partial trace over the environmental degrees of freedom and evolves deterministically under a quantum master equation. This approach requires starting from a microscopically explicit model of the environment and its interactions with the system, followed by either numerically-intensive reduction schemes, such as the hierarchical equations of motion (HEOM) or the time-evolving density matrix using the orthogonal polynomials algorithm (TEDOPA) \cite{Tanimura1989TimeBath, Tanimura1990NonperturbativeBath,Tanimura2020NumericallyHEOM,Werner2016PositiveSystems,Lacroix2024MPSDynamics.jl:Dynamics,Cygorek2022SimulationEnvironments},
or clear-cut approximations on the interaction strength and timescale separation, as in the seminal derivation of the reduced dynamics proposed by Redfield \cite{Redfield1957,Redfield1965TheProcesses}.

In the second approach, the density matrix is understood as the statistical average over an ensemble of stochastic pure-state trajectories. The latter viewpoint naturally motivates the introduction of stochastic Hamiltonians, in which the environmental influence is represented by time-dependent random fluctuations of system parameters.
Averaging over realizations recovers the mixed-state dynamics of an open system, although the environment degrees of freedom are never considered explicitly. This framework offers a physically intuitive and computationally versatile route to include environmental effects in the dynamics, and has proven particularly powerful in the interpretation of spectroscopic observables where noise-induced dephasing and spectral diffusion play a central role and are reflected in the spectral lineshape \cite{Anderson1954,Kubo1954AAbsorption,Kubo1963StochasticEquations,Haken1973AnMotion,Sanda2008StochasticExcitons}.
More recently, a wide range of sophisticated techniques have been developed to control and drive the quantum state for information processing. When the control field is described within a semiclassical framework, the inclusion of fluctuations in the driving protocol directly maps onto a stochastic Hamiltonian description. This perspective is especially relevant in the context of noisy intermediate-scale quantum (NISQ) devices \cite{Preskill2018QuantumBeyond,Brandhofer2021SpecialThem,Bharti2022NoisyAlgorithms}, where unavoidable imperfections and control noise manifest as stochastic perturbations of the system Hamiltonian, thereby shaping both dynamics and computational fidelity \cite{DeKeijzer2025QubitNoise,Baratz2025Data-DrivenDecomposition,Cialdi2017All-opticalChannels,Cialdi2019ExperimentalDynamics}.   

In this work, we start from the dynamics induced by a stochastic Hamiltonian \cite{Kubo1963StochasticEquations,Haken1973AnMotion,Hasegawa1980StochasticProcesses}, as it provides a particularly direct and versatile route to model environmental effects without the need for a complete microscopic model of the underlying system-environment dynamics. 
In this framework, the effective Hamiltonian $H^\mathrm{eff}_t$ is composed of the bare system Hamiltonian $H$ plus a randomly fluctuating term $H_t^\mathrm{fluct}$, so that we can write 
\begin{equation}
\label{eq:eff_stoch_hamiltonian}
    \hat{H}^\mathrm{eff}_t = \hat{H} + \hat{H}_t^\mathrm{fluct} = \hat{H} + Z_t\hat{R}
\end{equation}
where $\hat{R}$ is a Hermitian operator in the system state that defines the interaction with the environment, and $(Z_t)_{t\geq0}$ is either a continuous stochastic process or a discontinuous  noise that
models the effect of the environment on the system.
Since the specific form of the system Hamiltonian $H$
is not relevant to our discussion, we will consider it as a generic time-independent Hermitian operator, but our conclusions can be easily generalized to the time-dependent case.

Direct introduction of random fluctuations in the Hamiltonian is not the only way of introducing stochastic dynamics in the quantum evolution. The stochastic Schrödinger equation (SSE) is a more general formalism in which the pure state of an open quantum system evolves according to a stochastic differential equation driven by noise terms \cite{Gisin1984QuantumProcesses, Barchielli1991MeasurementsMechanics,Diosi1988ContinuousFormalism,Diosi1988LocalizedEquation,Gisin1992TheSystems,Carmichael1993An18}.
These noise terms were interpreted in connection with specific measurement protocols,  ``quantum jumps'' or continuous measurement back-action \cite{petruccione2002,Jacobs2006AMeasurement,Barchielli2009Quantum782}.
Ultimately, the SSE is constructed so that the ensemble average over trajectories reproduces the deterministic evolution dictated by a master equation (in Lindblad form or non-Markovian extensions thereof \cite{Barchielli2009Quantum782,Diosi1997TheSystems,Breuer1999StochasticEquations,Semina2014StochasticCases,Li2021MarkovianEquations,Gasbarri2018StochasticMaps}). 

Renewed interest in stochastic methods in quantum dynamics has recently emerged in diverse contexts, ranging from molecular modeling \cite{Coccia2018ProbingSystems,DallOsto2024StochasticSystems} to quantum technologies \cite{Huang2020QubitNoise,DeKeijzer2025QubitNoise}.
The stochastic formulation of the open-system dynamics replaces an explicit system–environment evolution with an effective system-only description, drastically reducing complexity. 
It also offers an efficient platform for the simulation of many-body systems (spin glasses, molecules, nanostructures), where computing the density-matrix evolution is often prohibitive due to the quadratic scaling of the state dimensions, compared to the linear scaling of the wavefunction.
However, the shift from a deterministic to a stochastic dynamics affects both the formal apparatus and the interpretive framework.

In this paper, we aim to provide a clear analysis of these methods by uncovering and elaborating on their common roots in the basic concepts of stochastic calculus. We wish to present a unified perspective that bridges methodological developments using stochastic Hamiltonians across chemical physics and quantum information science.
This work is not intended to provide a comprehensive review of all the stochastic approaches to open quantum systems; rather, it assembles selected tools and results known from specific applications and draws explicit connections among them within a single stochastic-calculus framework. Our main interest is in dynamics governed by stochastic Hamiltonians (SH), which we identify as a constrained instance of the stochastic Schrödinger equation (SSE) for pure states and of the stochastic Liouville equation (SLE) for generally mixed states. We trace how different choices in the formalism determine implementation pathways: numerical quadrature of stochastic trajectories versus random-unitary maps, the latter being well-suited for integration into quantum algorithms. In this sense, the contribution of this work is a self-contained synthesis that clarifies foundations, equivalences, and computational consequences, yielding a practical toolkit for simulating open-system dynamics with stochastic Hamiltonians.

To this end, we will make use of the two main interpretations of stochastic calculus, named after \ito and Stratonovich, presented in \Cref{sec:StochasticCalculusTools}, and point out explicitly the difference between a stochastic process and a noise. We further show how each formulation naturally maps to distinct numerical workflows, amenable to efficient classical trajectory integration or to quantum-algorithmic implementations.
\Cref{fig:StochasticHamiltonian_frameworks_scheme} illustrates graphically the results presented in \Cref{sec:stratonovich,sec:ito,sec:rode}. First, in \Cref{sec:stratonovich}, we discuss the lower blue pathway.
\Cref{eq:stoch_schr_eq_gen} is interpreted in the Stratonovich formalism as a viable way to implement quantum simulations, leading directly to the composition of universal gates in a quantum computing architecture.
In \Cref{sec:ito}, we follow the upper green branch of the schematic map and adopt the \ito interpretation,
showing that it is the correct framework to obtain valid quantum master equations (QME).
We also discuss how to write stochastic \schr equations (SSE), and clarify how a stochastic Hamiltonian model is a special case of the SSE subject to more stringent constraints.
In addition, we show that the \ito form of the evolution equations is easily treated with classical efficient time-integration schemes.
A particular class of equations that can be cast indifferently in both \ito and Stratonovich frameworks is briefly discussed in \Cref{sec:rode}. 
Finally, in \Cref{sec:sle}, we show how we can work directly with stochastic density matrices instead of wavefunctions, moving back to the framework of earlier seminal works that employed stochastic Liouville equations \cite{Kubo1963StochasticEquations,Haken1973AnMotion}.
This allows for the inclusion of various effects, such as different dissipators and measurement apparatuses.  

\begin{figure*}
    \centering
    \begin{tikzpicture}[node distance=2cm, thick, >=Stealth]
        \tikzstyle{block} = [draw, rounded corners, minimum width=2.2cm, minimum height=1.2cm, align=center]
        \tikzstyle{circleblock} = [draw, circle, minimum width=2.2cm, align=center]
        \node (start) {$\hat H^{\text{fluct}}_t \psi_t^{\text{traj}}$}; 
        \deepgreen{
        \node[block, right=3.5cm of start] (sse) { {\Large SSE} \\ Stoch. Schr. Eq.}; 
        \node[block, right=4cm of sse] (master) {$\dfrac{d\rho_t}{dt}$};
        \node[right=3mm of master] {QME}; 
        \draw[->,above] (start) to node {$\psi^\mathrm{trj}\mathrm{d}W_t$} (sse);
        \draw[below] (start) to node {It\={o}} (sse);
        \draw[->,above] (sse)  to node {$\mathbb{E}_\Omega\Big[ \mathrm{d}\left(|\psi_t^\mathrm{trj}\rangle\langle\psi_t^\mathrm{trj}| \right) \Big]$} (master);
        \draw[below](sse)  to node {\textit{average}} (master);
            \node[right=3.5cm of start] (tmpforbelow) {};
        }
        \deepblue{
        \node[circleblock, below=1.5cm of sse] (ut) {{\Large 
        $U_t$} \\ Random \\ Unitary \\ Gates}; 
        \node[circleblock,below=1.5cm of master] (rho) {$\rho_t$};
            \node[below=2.6cm of tmpforbelow] (strato) {};
            \node[below=3.8cm of tmpforbelow] (rode) {};
        \draw[->] (start)+(3mm,-5mm) |- (strato) 
        node[midway, above right, xshift={+12mm}] {$\psi^\mathrm{trj}\circ\mathrm{d}W_t$}
        node[midway, below right, xshift={+7mm}] {Stratonovich};       
        \draw[->] (start)+(-3mm,-5mm) |- (rode)
        node[midway, above right, xshift={+15mm}] {$H\in\mathcal{C}^0$}
        node[midway, below right, xshift={+15mm}] {RODE};
        \draw[->,above] (ut) to node {$\mathbb{E}\Big[ \mathcal{U}_t|\psi_0\rangle\langle\psi_0|\mathcal{U}_t^{\dagger}  \Big]$} (rho);
        \draw[below] (ut) to node {\textit{average}} (rho);}
        \draw[->] (master) -- (rho);
    \end{tikzpicture}
    \caption{A schematic map of the two mathematical frameworks and relative workflows for the stochastic Hamiltonian applied to a wavefunction dynamics, based on the \ito (upper green branch) and Stratonovich (lower blue branch) interpretations.
    The Stratonovich pathway connects stochastic Hamiltonians to direct simulations of the density matrix dynamics through random unitary gates, while the \ito pathway provides the correct formulation for analytically obtaining quantum master equations (QME).
    The random ordinary differential equation (RODE) case appears as a special limit when dealing with continuous stochastic processes as Hamiltonian fluctuations.}
    \label{fig:StochasticHamiltonian_frameworks_scheme}
\end{figure*}

\section{Physical context and mathematical background}\label{sec:physbakcground_mathtools}

We now introduce the physical settings and mathematical frameworks on which our discussion builds. First, we describe how stochastic Hamiltonians arise from system–environment interactions and how they are modeled effectively within the system’s Hilbert space. Then, we review the key elements of stochastic calculus, with emphasis on the \ito and Stratonovich interpretations of stochastic differentials, which will play a central role in the developments of the subsequent sections and provide the mathematical language for the formulations that follow, as well as the definitions of stochastic process and noise.

\subsection{Modeling of the stochastic Hamiltonian}\label{sec:modeling_SH}
We start from a system-environment interaction Hamiltonian of the type  
\begin{equation}
    \hat{H}_\text{int} = \left(c \hat{S} + c^* \hat{S}^\dagger\right) \hat{B}    
\end{equation}
where $\hat{S}$ acts on the system and $\hat{B}$ is a Hermitian operator acting on the environment. 
To confine the description within the Hilbert space of the system only, the environment operator $\hat{B}$ must be replaced with a function $B_t$. This is the first step of numerous embedding schemes, where the environment is reflected in a possibly time-dependent potential. 
Notice that when such a potential has a stochastic character, the expected change of the purity of the system state, typical of open quantum systems, can be recovered by averaging over different trajectories \cite{Kubo1966TheTheorem,Gaspard1999Non-MarkovianEquation}. 

In some cases, the stochastic field can be considered as a surrogate environment, originating from the coupling Hamiltonian and the state of the original microscopic environment \cite{Szankowski2020NoiseDynamics,Szankowski2021MeasuringNoise,Gu2019WhenNoise}.
A prominent example used in triplet exciton dynamics is the replacement of a wide-bandwidth bosonic bath in the high-temperature limit with Gaussian white noise, which acts as a genuine surrogate field \cite{Haken1973AnMotion}.
More commonly, the use of stochastic fluctuations in the Hamiltonian is postulated to represent the dominant environmental influence, with no reference to a microscopically defined environment. 
This phenomenological approach is conceptually different from the exact treatment of open quantum dynamics, where the elimination of the environment’s degrees of freedom leads to specific stochastic forms like those originating 
from the Feynman-Vernon (path-integral) formalism \cite{Feynman1963TheSystem,Grabert1988QuantumApproach}, non-Markovian quantum state diffusion (NMQSD) \cite{Diosi1997TheSystems,Diosi1998Non-MarkovianDiffusion,Stockburger2001Non-MarkovianDiffusion}, the hierarchy of pure states (HOPS) \cite{Suess2014HierarchyDynamics}, the HEOM approach\cite{Tanimura1989TimeBath, Tanimura1990NonperturbativeBath,Tanimura2020NumericallyHEOM}, the generalized Langevin equations (GLE) \cite{Stella2014GeneralizedSystems,Ness2015ApplicationsBaths,Matos2020EfficientSystems} and others \cite{Stockburger2002ExactDissipation,McCaul2017Partition-freeEquation,Magazzu2022Feynman-VernonStudy}.

The fluctuating interaction Hamiltonian, now acting on the system subspace only, becomes
\begin{equation}
\hat{H}_t^\mathrm{fluct} = Z_t \hat{S} + Z_t^* \hat{S}^\dagger    
\end{equation}
where 
\begin{equation}
    Z_t = c B_t.
\end{equation} 
For real interaction coefficients $c$, we obtain real processes $Z_t$, allowing us to identify the operator
\begin{equation}
\label{eq:hermitian_R}
    \hat{R} = \hat{S} + \hat{S}^\dagger,
\end{equation}
and retrieve \cref{eq:eff_stoch_hamiltonian}.

The dynamics of the reduced system, which originates from the evolution of the system plus environment under the Hamiltonian $\hat{H}' = \hat{H} + \hat{H}_\text{int}$, unravels in terms of trajectories.
We can write the \schr equation for a system with Hamiltonian given by \cref{eq:eff_stoch_hamiltonian} as
\begin{equation}
\label{eq:stoch_schr_eq_gen}
    \frac{\diff}{\diff{t}}{\psi_t^\mathrm{trj}} = -i\left(\hat{H} + Z_t\hat{R} \right){\psi_t^\mathrm{trj}},
\end{equation}
where the superscript acknowledges the random nature of the single evolution, a trajectory. 
Depending on the formal structure of $Z_t$ and the target implementation, different formalisms should be used to ensure that \cref{eq:stoch_schr_eq_gen} is well defined,
as we show below. 
From now on, to avoid cluttering the notation, we drop the hat symbol on the operators where the nature of each term is clear from the context.

The reduced density matrix of the open system is obtained as the averaged density matrix over all the possible trajectories, that is
\begin{equation}
\label{eq:rho_as_average}
    \rho_t = \mathbb{E}\left[\rho^\mathrm{trj}_t\right] =\mathbb{E}\left[\ketbra{\psi_t^\mathrm{trj}}{\psi_t^\mathrm{trj}}\right],
\end{equation}
where $\rho^\mathrm{trj}_t$ is the density matrix of the system along a trajectory
and $\mathbb{E}[\cdot]$ indicates the expected value of the argument, obtained as the ensemble average.
In any practical case, the average over the ensemble of all possible trajectories $\omega$ is approximated by a finite averaging
\begin{equation}
\label{eq:average}
    \mathbb{E}[\rho^\mathrm{trj}_t] =
    \lim_{n\to\infty}\frac{1}{n}\sum^n_{\omega=1}\rho_t^{(\omega)} \
    \simeq
    \frac{1}{N}
    \sum^N_{\omega=1}
    \rho_t^{(\omega)} \,.
\end{equation}

We remark that the averaged density matrix, \cref{eq:rho_as_average}, does not describe a pure state, but rather the mixed state of an ensemble where dissipative effects stem from the fluctuations induced by the stochastic component.
Importantly, the average maps are completely positive and trace-preserving (CPT) \cite{Gasbarri2018StochasticMaps} by construction, as the mean dynamics derives from the average of pure quantum states. 

\begin{figure}
    \centering
    \begin{tikzpicture}[>=Stealth, thick, scale=1.]
        \definecolor{darkred}{RGB}{180,30,30}
        \definecolor{bluearrow}{RGB}{0,150,255}
            \def\xshiftzero{-1.5} 
            \def\xshiftone{1.5}
            \def\centerdotspos{2}
            \def\yshiftzero{2.3}
            \def\yshiftone{1.6}
            \def\yshiftbottoms{-0.5}
            \def\halfarrowslength{1.}
            \def\deltayshiftpos{.5}
            \draw[darkred, line width=1.5pt] (\xshiftzero,\yshiftzero) -- (\xshiftzero+1,\yshiftzero);
            \draw[darkred, line width=1.5pt] (\xshiftone,\yshiftone) -- (\xshiftone+1,\yshiftone);
            \draw[dotted, very thick] (\xshiftzero,\centerdotspos) -- (\xshiftone+1,\centerdotspos);
            \draw[thick, ->] (3.5,\centerdotspos-1) -- (3.5,\centerdotspos+1);
            \node[left=2pt] at (3.5,1.) {$E$};
           
            \draw[bluearrow, <->, line width=1.2pt] (\xshiftzero+0.5,\centerdotspos+\halfarrowslength) -- (\xshiftzero+0.5,\centerdotspos-\halfarrowslength);
            \draw[bluearrow, <->, line width=1.2pt] (\xshiftone+0.5,\centerdotspos+\halfarrowslength) -- (\xshiftone+0.5,\centerdotspos-\halfarrowslength);
            
            \node[above=1pt] at (\xshiftzero+0.5,\yshiftzero+\deltayshiftpos) {};
            \node[below=1pt] at (\xshiftone+0.5,\yshiftone-\deltayshiftpos) {};
            
            \node[left=2pt] at (\xshiftzero,\yshiftzero) {$\varepsilon_0+\delta\varepsilon_0(t)$};
            \node[left=2pt] at (\xshiftone,\yshiftone) {$\varepsilon_1+\delta\varepsilon_1(t)$};
        \draw[<->,bend left=50] (\xshiftzero+1.05,\yshiftzero) to  node[align=right,xshift=2,yshift=5]{$\Omega$} (\xshiftone-0.05,\yshiftone);
    \end{tikzpicture}
    \caption{Example of the effect of a stochastic fluctuation at a time $t$ on the Hamiltonian energy levels in site basis for the first exciton manifold of a system of two degenerate sites, i.e.\ $\varepsilon_0=\varepsilon_1$, indicated by the central dotted line.
    The vertical blue lines indicate the variance of the stochastic energy fluctuation, the red lines the instantaneous energy of each level at a time $t$, namely $\varepsilon_i+\delta\varepsilon_i(t)$.
    The two levels are coupled by a term of constant $\Omega$, shown as a black line.}
    \label{fig:shochFluctuationEnergyLevels}
\end{figure}

To exemplify the formalisms introduced above and the further development in the following sections, we will consider examples of quantum two-level systems. In what follows, we recall the stochastic Hamiltonian formulation of coherence decay between two excited states, originally proposed in the context of triplet exciton dynamics by Haken and Strobl \cite{Haken1973AnMotion}.

\begin{figure}
    \centering
    \includegraphics[width=\linewidth]{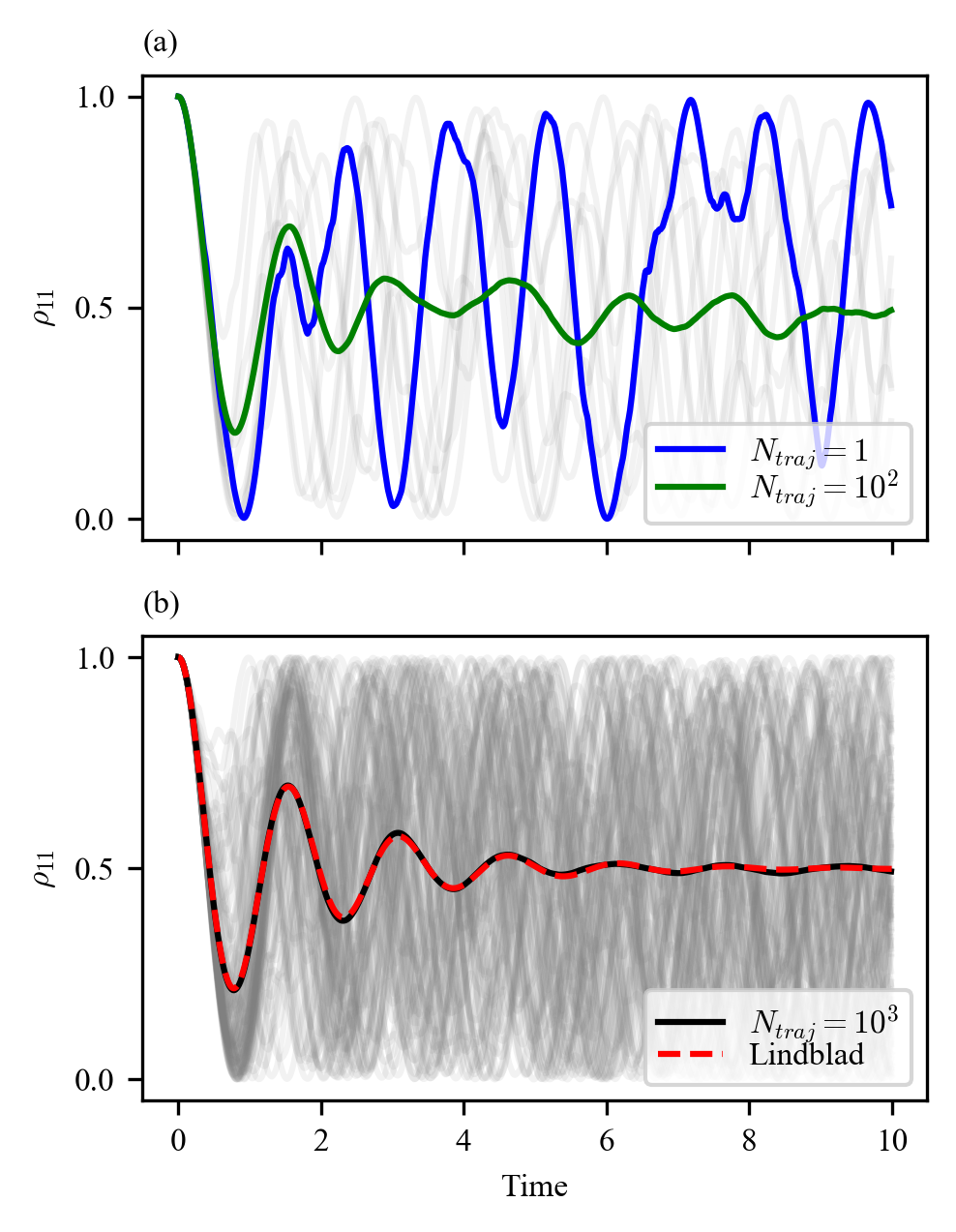}
    \caption{Stochastic trajectories and average dynamics of the excited-state population $\rho^{(11)}_t = \bra{1} \rho_t \ket{1}$ for the Haken–Strobl model presented in the text, computed with the stochastic Hamiltonian approach.
    (a) Individual realizations of a small ensemble of $10^2$ trajectories (grey lines in transparency), one single trajectory highlighted (blue line), and their average (green line).
    In panel (b), the average over $10^3$ trajectories (black line) is compared with the exact solution of the corresponding Lindblad master equation (red dashed line), showing agreement between the stochastic and the master equation descriptions.}
    \label{fig:hakenstrobl_SH_dynamics}
\end{figure}

\vspace{1ex}
\textbf{Example 1 (Haken-Strobl) }
Let the system basis be composed of two sites $\{\ket{0},\ket{1}\}$, with energy $\varepsilon_0$ and $\varepsilon_1$, respectively, and let them interact by 
a coupling $\Omega$ that coherently switches between the sites.
The total Hamiltonian, $H$ in eq. \eqref{eq:eff_stoch_hamiltonian}, is then
\begin{equation}
    H = \varepsilon_0\ketbra{0}{0}+\varepsilon_1\ketbra{1}{1} + \Omega \big(\ketbra{0}{1}+\ketbra{1}{0}\big) \,.
\end{equation}
A stochastic fluctuation, modeled as white noise (see \Cref{app:whitenoise}),
is applied to the site energies to replicate the effect of a rapid Markovian environment, \cref{fig:shochFluctuationEnergyLevels}.
We can therefore write the fluctuation Hamiltonian as 
\begin{equation}
    \hat{H}_t^\mathrm{fluct} = \delta\varepsilon_0(t) \ketbra{0}{0} + \delta\varepsilon_1(t) \ketbra{1}{1} \,.
\end{equation}
From a comparison with eq. \eqref{eq:eff_stoch_hamiltonian}, we can identify two noise variables $Z_0(t) = \delta\varepsilon_0(t)$ and $Z_1(t) = \delta\varepsilon_1(t)$, and the two associated operators $R_0 = \ketbra{0}{0}$ and $R_1 = \ketbra{1}{1}$.
This setting reflects a Haken-Strobl model \cite{Haken1973AnMotion}, the prototypical description of site dephasing in a quantum system.
The trajectory dynamics are computed according to \cref{eq:stoch_schr_eq_gen}, the formal explanation for this choice will become clear in \Cref{sec:stratonovich}, and the mean dynamics is obtained as the average over trajectories, see \cref{eq:average}. It is well known (see \Cref{sec:ito} for the details) that the average dynamics of this model is exactly the solution of a Lindblad equation, reported in \Cref{app:HS_Lindblad}.
In \cref{fig:hakenstrobl_SH_dynamics}, a swarm of trajectories is depicted, showing the random character of each realization, and the effect of the different sizes of the sample space is highlighted in the two panels by comparison with the exact solution of the associated Lindblad master equation. 
Note that, differently from the average behavior, each trajectory maintains the coherent (oscillating) character typical of pure states.
The system parameters used for the example are reported in \Cref{app:HS_Lindblad}.

\subsection{Stochastic calculus tools}
\label{sec:StochasticCalculusTools}

Introducing stochastic potentials to model random contributions to a system's dynamics naturally calls for the use of stochastic calculus tools. 
Within this section, we provide a brief review of essential concepts. We recall the differences between the approaches by \ito \cite{Ito1944StochasticIntegral} and Stratonovich \cite{Stratonovich1966AEquations} to stochastic calculus as they will be central to appreciate the application-oriented analysis of the following sections. 

We introduce the definitions of stochastic differential and stochastic integral, 
needed to formalize the definitions of stochastic process and noise, in the two main frameworks that we need in the following.
For a comprehensive and accessible introduction to the topic, we refer the reader to Ref.\ \cite{Evans2013AnEquations}.

We define a stochastic differential equation (SDE) as the formal expression
\begin{equation}
\label{eq:gen_SDE}
    \diff{X_t} = f(t,X_t)\diff{t} + g(t,X_t)\diff{W_t} 
\end{equation}
where $\diff{X_t}$ is the stochastic differential of the process $X_t$,
$f(t,X_t)$ and $g(t,X_t)$ are either functionals of the process or simple functions of time,  $W_t$ is a Brownian motion (Wiener process), and $\diff{W_t}$ its differential with scaling $\diff{W_t}\sim \sqrt{\diff{t}}$, see \cref{app:whitenoise} for more details.
Indeed, this type of notation is needed because the time derivative of the Wiener process is not a well-defined function; hence, the same is true for the time derivative of the generic process $X_t$.

Eq. \ref{eq:gen_SDE} is a shorthand notation for the integral equation
\begin{equation}
\label{eq:integral_SDE}
    {X_t} = X_0 + \int_0^t f(s,X_s)\diff{s} + \int_0^t g(s,X_s)\diff{W_s} ,
\end{equation}
implying that we need to find a way to define an integral of the form 
\begin{equation}
    \int_0^T g(t,X_t) \diff{W_t},
\end{equation}
that cannot be understood as an ordinary integral. 

Expressing the integral as its Riemann-sum approximation on a partition of $m$ elements of the time interval $[0,T]$,
we write
\begin{equation}
\label{eq:riemannsum}
    \int_0^T g(t,X_t) \diff{W_t} \approx
    \sum_{k=0}^{m-1} g(\tau_k, X_{\tau_k})  \left( W(t_{k+1}) - W(t_k) \right)
\end{equation}
where $\tau_k$ is a point that lies within the subinterval
$[t_k,t_{k+1}]$, i.e.
\begin{equation}
    \tau_k = \lambda t_{k+1} +(1-\lambda) t_{k}
    \quad
    \lambda\in[0,1].
\end{equation}
The choice of the parameter $\lambda$ defines different integral interpretations, introducing different frameworks \cite{Evans2013AnEquations,Ito1975StochasticCalculus}.

\ito's integration is identified by $\lambda \!=\! 0$, which corresponds to setting 
$\tau_k \!= t_k$, the left-hand point of each subinterval.
This choice means that we do not need previous knowledge of the function at future values,
in agreement with the idea that the stochastic potential provides unpredictable random perturbations to the system.
In this interpretation of the stochastic differentials, the product rule for the differentials is
not the ordinary one, but
\ito's product rule:
for two differentials of the form $\diff X_i = F_i \diff t + G_i \diff W_t$, the differential of the product
is
    \begin{equation}
        \diff(X_1 X_2) = X_2 \diff X_1 + X_1\diff X_2 + G_1 G_2\diff t,
        \label{eq:ItoProduct}
    \end{equation}
where, compared to ordinary calculus, we find the additional term $G_1 G_2\diff t$ due to the scaling $(\diff{W})^2\sim\diff{t}$.

Alternatively, among the infinite other possible choices,
setting $\lambda=\frac{1}{2}$ leads to the Stratonovich integral definition.
In this definition, the integral is time-symmetric, 
with both the function and the Wiener differential ``sticking out toward the future''
and the functional $g(X_t,t)$ being now \textit{anticipating}, or \textit{non-adapted}.
This means that the information at any time $t$ depends on some future-time values, but it agrees with the notion of a mid-point rule.
Thus, in the differential Stratonovich notation, commonly identified by
a circle before the stochastic differential $\circ\diff{W}$, we can write
\begin{equation}
\label{eq:stratonovich_differentialproduct}
    \circ\diff (X_1 X_2) = X_1\circ\diff{X_2} + X_2\circ\diff{X_1}
\end{equation}
where the usefulness of restoring the basic rules of differential calculus justifies the tricky anticipatory nature of this interpretation.
A fundamental point now is the definition of
the $\circ$ symbol we used to indicate the Stratonovich differential as an actual multiplication, known as 
the \textit{symmetric Q-multiplication} \cite{Ito1975StochasticCalculus} which,
for two time-dependent random variables $X_t$ and $Y_t$, 
is
\begin{equation}
\label{eq:stratonovich_product}
    Y_t\circ\diff X_t := Y_t\diff{X_t} +\frac{1}{2}\diff X_t\diff{Y_t}
\end{equation}
with the differential on the r.h.s. now expressed in the \ito interpretation.
Then, by combinations, and possibly iteration, of \cref{eq:stratonovich_differentialproduct} we can transform a Stratonovich SDE into an \ito SDE.

Hereforth, we define a stochastic \textit{process} as the integrated form of any \ito stochastic differential equation,
$X=\int\diff{X_t}$, that is the solution of \cref{eq:integral_SDE}.
The term \textit{noise} corresponds instead to the \textit{formal} time derivative of the stochastic process in a distributional sense, $\dot{X_t}=\diff{X_t}/\diff{t}$.
Therefore, in the SDE formulation, we define \textit{noise-inducing} the stochastic differential $\diff{X_t}$ itself. 
This definition can be understood by formally dividing \cref{eq:gen_SDE} by the time differential $\diff{t}$ and obtaining a Langevin-like form, where the presence of a noisy drive is evident in the equation of motion for the process.

\section{Unitary propagation under stochastic Hamiltonians}
\label{sec:stratonovich}

Looking for the solution of \cref{eq:stoch_schr_eq_gen}, under ordinary calculus rules, we expect to be able to write a formal propagator of the form
\begin{equation}
    \label{eq:unitary_propagator}
    \mathcal{U}_{t,t_0} = \timeorder \exp\left\{ -i \int_{t_0}^t H^\mathrm{eff}_s  \diff{s}   \right\} 
\end{equation}
where $\timeorder$ is the time-ordering operator, leading to solutions of the wavefunction evolution in the form  
\begin{equation}
    \psi_t = \mathcal{U}_{t,t_0} \psi_0
\end{equation}
with initial value $\psi_0$, meaning that we should be able to propagate each trajectory through this unitary.

As anticipated in the introduction, we want to write \cref{eq:stoch_schr_eq_gen} in a formally correct way.
We start writing the random potential as a noise,
for simplicity the common white noise $\xi_t$ of intensity $\gamma$, by
setting
\begin{equation}
    Z_t = \xi_t
    = 
    \gamma\frac{\diff{W_t}}{\diff{t}},
\end{equation}
where $\diff{W_t}$ is a proper stochastic differential and the formal time derivative is a distribution, the white noise distribution.
Substituting this definition into \cref{eq:stoch_schr_eq_gen} and multiplying both sides by $\diff{t}$, we obtain a mathematically correct formulation of the equation:
\begin{equation}
\label{eq:Stratonovich_SSE}
    \diff\psi_t = -i{H}\psi_t\diff{t} -i \gamma {R} \psi_t \circ\diff{W_t}.
\end{equation}

The choice of the Stratonovich interpretation of the differential is fundamental, as it allows us 
to write the analytical solution to \cref{eq:Stratonovich_SSE}, 
since in this formalism the rules of ordinary calculus are indeed valid. 
As for the solution of the \schr equation, we obtain
\begin{equation}
\label{eq:stratonovich_solution_white}
    \psi_t 
    = \timeorder \exp\left\{ -i \int_0^t H\diff{t} - i \gamma \int_0^t R\circ\diff{W_t}  \right\} \psi_0 \,,
\end{equation}
where we recognize in the time-ordered exponential the propagator of the dynamics,
meaning that each trajectory can be propagated through this unitary.
These solutions,
proposed using white noise for the sake of clarity, can be easily generalized for colored noise, paying attention to the correct integration of the deterministic and stochastic parts.
We point out, in the following \Cref{sec:ito}, that this is not always the case in the \ito formalism, and further constraints must be introduced.
 
Care must be used in the description of the dynamics of the mean density matrix of the open system, whose equation cannot be obtained by adhering to this formalism.
Indeed, while \cref{eq:rho_as_average} holds true,
if we take the product differential $\diff{(\psi_t\psi_t\daggah)}$ and we average,
\begin{equation}
    \mathbb{E}\left[\diff{(\psi_t\psi_t\daggah)}\right] = -i \left[H, \psi_t\psi_t\daggah\right]\diff{t} - i
    \gamma\,
    \mathbb{E}\left( \left[R, \psi_t \psi_t\daggah \right]\circ\diff{W_t}\right) 
\end{equation}
we do not directly obtain the expected form of the master equation, a Lindblad form in the case of white noise stochastic driving.
We refer to \Cref{sec:ito} for the computation of the QME, for which one needs to work with \ito calculus.

\subsection{Use of unitary propagators for implementation of quantum algorithms}
\label{sec:unitary_quantumalg}

In the framework discussed above,  we obtain the expression of the propagator as a unitary operator, and norm-preservation is thus ensured trajectory-wise.  
We can obtain the numerical solution for the evolution in the time interval $[0,T]$, discretizing time into $S$ small time steps $\tau=T/S$
\begin{equation}
\label{eq:propagator_discretization}
    \mathcal{U}_{t} =\ \prod_{s=1}^S 
\ \mathcal{U}_{s\tau,(s-1)\tau}
\end{equation}
and approximating at each time step the time-ordered unitary propagator with a first-order Magnus expansion \cite{Magnus1954OnOperator,Blanes2009TheApplications}:
\begin{equation}
\begin{split}
    \mathcal{U}_{s\tau,(s-1)\tau} \approx \exp\Bigg\{ &-i \int_{(s-1)\tau}^{s\tau} H \diff{r}     \\
    &-i \gamma\int_{(s-1)\tau}^{s\tau} R \circ\diff{W_r}   \Bigg\}.
\end{split}
\end{equation}
This allows for the numerical integration of every trajectory solution, keeping the unitary character of the propagators.
We highlight that the choice of a small time step is important for convergence to the exact time-ordering solution.
Higher-order Magnus expansions can also be employed to achieve better convergence.

For a time-independent system Hamiltonian and noise operator, the short-time evolution operators read
\begin{equation}
\label{eq:propagatore_per_esempio_QA}
        \mathcal{U}_{\tau} = \exp\left\{ -i H \tau -i \gamma R \Delta{W_\tau}  \right\},
\end{equation}
where the white-noise increment $\Delta{W_\tau}$ is drawn from a normal distribution of zero mean and standard deviation the square root of the time interval:
\begin{equation}
    \Delta{W\tau} \sim \mathcal{N}(0,\tau)=\sqrt{\tau}\,\mathcal{N}(0,1).
\end{equation}

This numerical scheme, also used for implementations on classical computers \cite{Jansen2006, Liang2012}, allows the construction of quantum algorithms for digital quantum computers without the overhead of auxiliary resources (auxiliary qubits).
This is because the unitary character ensured by the use of the Magnus expansion is naturally suited to be implemented in the language of quantum gates, the unitary operations that represent the building blocks of every quantum circuit \cite{Nielsen2010QuantumEdition}.

\vspace{1ex}
\textbf{Example 2 (Quantum Circuit) }
A simple example of a quantum algorithm \cite{gallina2022,Gallina2024FromDynamics} that implements eqs. \eqref{eq:propagator_discretization} and \eqref{eq:propagatore_per_esempio_QA}
 is depicted in \cref{fig:circuit_QCNA}.
The quantum circuit reproduces the evolution of a two-level system (qubit) with states $\ket{0}$ and $\ket{1}$) dictated by a system Hamiltonian 
$H = \varepsilon\sigma_z + \Omega\sigma_x$, where $\sigma_z = \ketbra{0}{0}-\ketbra{1}{1}$ and $\sigma_x = \ketbra{0}{1} + \ketbra{1}{0}$, and a fluctuation Hamiltonian $\hat{H}_t^\mathrm{fluct} = \xi_t \sigma_z$.
This is formally equivalent to the model used to represent the coupled excited states in \cref{sec:modeling_SH}, cast in terms of Pauli operators acting on a qubit. Explicitly, the energy difference of the sites is expressed by $\varepsilon=(\varepsilon_0 - \varepsilon_1)/2$ and so the fluctuation $\xi_t  = (\delta\varepsilon_0(t) - \delta\varepsilon_1(t)) / 2$.
The evolution operators, described in eq. \eqref{eq:propagatore_per_esempio_QA}, can be implemented in terms of quantum gates (involving 1 or 2 qubits) in various ways.
Here, for illustrative purposes, we show the notorious first-order Suzuki-Trotter decomposition \cite{Trotter1959OnOperators,Suzuki1976GeneralizedProblems}, which allows us to write
\begin{equation}
\label{eq:Trotterizzazione}
\begin{split}
    &\mathcal{U}_{\tau} = \exp\{-i \varepsilon \tau \sigma_z -i \Omega \tau \sigma_x -i \gamma \Delta W_\tau \sigma_z \}\\ &\approx \left[\exp\left\{-i \frac{\gamma \Delta W_\tau}{n} \sigma_z \right\} \exp\left\{-i \frac{\Omega \tau}{n} \sigma_x \right\} \exp\left\{-i \frac{\varepsilon \tau}{n} \sigma_z \right\}\right]^n
\end{split}
\end{equation}
where, in general, the larger $n$ the better the convergence.
However, due to the choice of a small time step dictated by the Magnus expansion, good results can be obtained even with a small $n$ (e.g., $n=1$).
Each unitary operator inside the squared brackets in eq. \eqref{eq:Trotterizzazione} is represented by a gate $R_\alpha(2 \theta) = \exp(-i \theta \sigma_\alpha)$, with $\alpha\in\{Z,X\}$, inside the red dashed barriers in \cref{fig:circuit_QCNA}.
To compute the dynamics of the system up to time $T = S\tau$, the 3-gate sequence is repeated $nS$ times. At the end, the qubit ($q_0$) is measured to know the population of the two states, $\ket{0}$ and $\ket{1}$.

\begin{figure}
    \centering
    \begin{quantikz}[scale=0.8]
        \lstick{$q_0:$} \slice{} &  \gate{R_Z(\theta_\varepsilon)}& \gate{R_X(\theta_\Omega)}& \gate{R_Z(\theta_\xi)}\slice{\#times} &  \meter{}  \wire[d][1]{b}    \\
        \lstick{$c$} \setwiretype{c}  & & & & &
    \end{quantikz}
    \caption{An example of a quantum circuit for the propagation of stochastic dynamics.
    Single qubits gates implement the unitary operators $R_\alpha(2 \theta) = \exp(-i \theta \sigma_\alpha)$, with $\alpha\in\{Z,X\}$, where $\sigma_\alpha$ are Pauli matrices.
    While parameters $\theta_\varepsilon$ and $\theta_\Omega$ are deterministic, $\theta_\xi$ is intended as a stochastic term changing at each time step and each repetition of the circuit.
    The Magnus-Trotter block evolving the system is performed multiple times to evolve the system up to the measurement.}
    \label{fig:circuit_QCNA}
\end{figure}

\section{Deriving master equations from stochastic dynamics}\label{sec:ito}

In this section, we present two different routes to obtain the equation that regulates the average dynamics. 
First, we show that the definition of stochastic \schr equations (SSE) is based on \ito's calculus.
We start with a general way to formulate the SSE, without delving into details out of our scope, but
pointing out along the way the more general descriptions that this method allows for.
We then rewrite the Stratonovich SDE in its equivalent \ito's form, the way we need to go to recover the Lindblad master equation, therefore showing the equivalence of the two methods if we impose constraints on the SSE.

A simple and general way to formulate the SSE is with a linear stochastic differential equation 
\begin{equation}
\label{eq:Ito_SSE_GENERAL}
    \diff\psi_t = A\psi_t\diff{t} + B\psi_t\diff{X_t}
\end{equation}
where the operator $A$ contains the deterministic (Hamiltonian)
evolution of the system and $B$ is the operator setting the structure of the stochastic fluctuations encoded by the \ito's differential $\diff X_t$ of a stochastic process $X_t$.
This allows for a great variety of models depending on the choice of the process and of the operators \cite{Barchielli2010StochasticNoise,Cialdi2019ExperimentalDynamics,DeChecchi2025DynamicsEquations,luczka2005Non-MarkovianNoise,DeKeijzer2025QubitNoise}.

Linear SSEs do not, in principle, ensure the preservation of the norm of the state vector during a trajectory. 
In this framework, normalization can be obtained by ensuring the martingale property on the norm of the wavefunction,
meaning that the norm of the wavefunction at any time is constant as an average over the ensemble.
This gives the condition
\begin{equation}
\label{eq:martingale_normaliz_cond}
    \mathbb{E}\left[||\psi_t||^2\right] =\mathbb{E}\left[||\psi_0||^2\right]=1 
    \implies     \mathbb{E}\left[\diff\big(\psi_t\daggah\psi_t\big)\right]=0 .
\end{equation}
Recalling \ito's product rule, \cref{eq:ItoProduct}, and writing in braket notation ($\psi=\ket{\psi},\,\psi\daggah=\bra{\psi}$), the condition above translates to
\begin{equation}
\label{eq:conditionAB}
\begin{split}
    \bra{\psi_t}&(A\daggah+A)\ket{\psi_t}\diff t +
    \bra{\psi_t}(B\daggah+B)\ket{\psi_t}\diff X_t \\[.5ex]
    +&\bra{\psi_t}B\daggah B\ket{\psi_t}(\diff X_t)^2 +
    \mathcal{O}(\diff t ^2) + \mathcal{O}(\diff t \diff X_t) = 0,
\end{split}
\end{equation}
where the second-order term w.r.t. $\diff X_t$
depends on the specific process $(X_t)_{\geq0}$ and terms of order $\mathcal{O}(t^{3/2})$ and higher are neglected. We note explicitly that these definitions do not make any assumptions on the stochastic process. Therefore, \Cref{eq:conditionAB} imposes constraints on the forms of operators $A$ and $B$ depending on the SDE of the noise $\diff{X_t}$.

\subsection{Averaging to the Lindblad form}

We now focus the discussion on white-noise-driven systems for the sake of clarity and for a later comparison with the Stratonovich SDE. 
We let the generic noise of the SSE in \cref{eq:Ito_SSE_GENERAL} be a white noise by setting
\begin{equation}
\label{eq:genericWhiteSSE}
    \diff\psi_t = A\psi_t\diff t + \gamma B\psi_t\diff W_t.
\end{equation}
When applying the normalization condition,
the term proportional to the noise (second term in \cref{eq:conditionAB}) averages to zero due to the properties of $\diff W$, see \cref{app:whitenoise} and ref.\ \cite{Evans2013AnEquations}.
Therefore, to preserve the martingale property, see \cref{eq:martingale_normaliz_cond,eq:conditionAB}, the following condition between the different operators must hold
\begin{equation}
\label{eq:normal_A}
    A\daggah + A + \gamma^2 B\daggah B = 0.
\end{equation}
This, together with the notion that the Hamiltonian must be contained in the operator of the deterministic part $A$,
leads to identifying it as the sum of the system Hamiltonian and a normalization term
\begin{equation}
\label{eq:Aoperatorwithcorrection}
    A = -{i} H -\frac{1}{2}\gamma^2 B\daggah B  ;  
\end{equation}
we stress that $(B\daggah B)$ is always Hermitian, 
even when $B$ is not. 
A step-by-step derivation of the condition of \cref{eq:normal_A} is presented in \cref{app:normalizationSSE}.

The normalized SSE is then
\begin{equation}
\label{eq:Ito_SSE_whitenorm}
    \diff\psi_t = \left(-iH -\frac{1}{2}\gamma^2 B\daggah B \right)\psi_t \diff t 
    +\gamma B \psi_t\diff W_t,
\end{equation}
which allows performing the numerical computation of the stochastic dynamics of the system state.
Using \ito's product rule, \cref{eq:ItoProduct}, for the differential $\diff{(\psi\psi\daggah)}=\diff{(\rho^\mathrm{trj})}$, we can write the corresponding equation for the evolution of the trajectory density matrix as
\begin{equation}
\begin{split}
    \diff(\rho^\mathrm{trj}_t) = 
    &-i[H,\rho^\mathrm{trj}_t]\diff t
    -\frac{1}{2}\gamma^2\left\{ B\daggah B, \rho^\mathrm{trj}_t \right\}\diff{t} \\[.5ex]
    &+ \gamma (B\rho^\mathrm{trj}_t + \rho^\mathrm{trj}_t B\daggah)\diff{W_t} + \gamma^2 B\rho^\mathrm{trj}_t B\daggah \diff{t}
\end{split}
\label{eq:QSME-OU-purestate}
\end{equation}
where $[\cdot,\cdot]$ is the commutator 
and $\{\cdot,\cdot\}$ the anti-commutator.
Finally, taking averages as in \cref{eq:rho_as_average}, we obtain the associated QME in the form
\begin{equation}
    \frac{\diff}{\diff{t}} \rho_t = 
    -i\left[H,\rho_t\right] 
    + {\Gamma}\left( B\rho_t B\daggah - \frac{1}{2}\left\{B\daggah B, \rho_t\right\} \right),
\end{equation}
which is a generic Lindblad form with a single relaxation channel characterized by a constant dissipation rate $\Gamma=\gamma^2$.
This is easily 
generalized to multiple dissipation channels using 
additive noise sources in \cref{eq:Ito_SSE_GENERAL}. 

Since there are no constraints on the operator $B$, this SSE describes a wide class of open system dynamics, wider than the stochastic Hamiltonian, as we clarify in the following section.
In particular, when $B$ does not commute with the Hamiltonian, the dissipative dynamics include transitions between the eigenstates of the Hamiltonian.
Furthermore, when $B$ is not Hermitian, upwards and downwards transitions in energy can occur with different rates, leading to non-uniform stationary states.

\begin{figure}
    \centering
    \includegraphics[width=\linewidth]{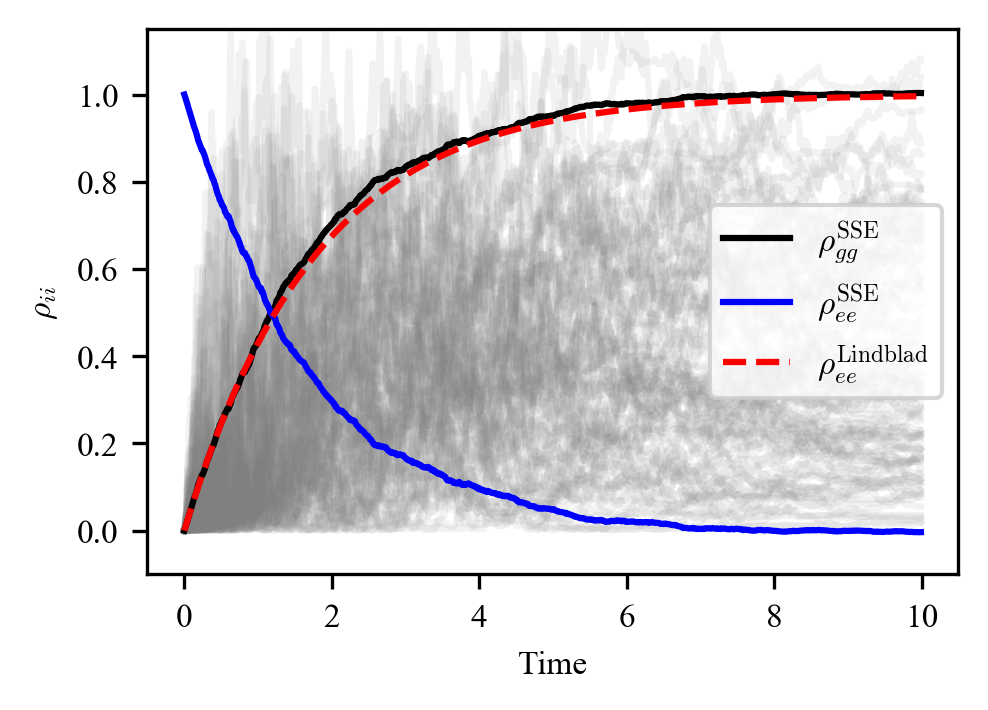}
    \caption{Spontaneous emission dynamics of a two-level system simulated via stochastic \schr equation (SSE) trajectories.
    The population of the excited state $\rho^{(ee)}_t$ (blue line) and the ground state $\rho^{(gg)}_t$ (black line) are obtained as averages over an ensemble of $10^3$ stochastic realizations, shown for $\rho^{(gg)}_t$ (grey transparency lines).
    The averages show agreement with the exact solution of the Lindblad master equation (red dashed line).}
    \label{fig:spontemiss_SSE_dynamics}
\end{figure}

\vspace{1ex}
\textbf{Example 3 (Spontaneous emission)} 
As an example of dissipative dynamics that can be described with a simple SSE but not with a stochastic Hamiltonian, we consider the spontaneous emission of a two-level system, from the excited state $\ket{e}$ to the ground state $\ket{g}$. The Hamiltonian of the system is  
\begin{equation}
    H = \varepsilon_g\ketbra{g}{g} + \varepsilon_e\ketbra{e}{e}\,.
\end{equation}
The operator for the stochastic component of the evolution is not Hermitian and takes the form $B=\ketbra{g}{e}$, also known as the $\sigma_-$ operator. This is the jump operator that, in the corresponding Lindblad dissipator, describes the transition from the excited to the ground state. 
The associated normalized SSE that we can propagate numerically is 
\begin{equation}
    \diff\psi_t = \left(-iH -\frac{1}{2}\gamma^2 \sigma_+\sigma_- \right)\psi_t \diff t 
    +\gamma \sigma_- \psi_t\diff W_t \,,
\end{equation}
where $\sigma_+=\sigma_-\daggah$, and the corresponding Lindblad equation for the average dynamics reads
\begin{equation}
    \frac{\diff}{\diff{t}} \rho_t = 
    -i\left[H,\rho_t\right] 
    + {\Gamma}\left( \sigma_-\rho_t \sigma_+ - \frac{1}{2}\left\{\sigma_+ \sigma_-, \rho_t\right\} \right).
\end{equation}
In terms of ground state population one obtains
\begin{equation}
    \frac{\diff}{\diff{t}} \rho_t^{(gg)} = \Gamma \rho_t^{(ee)}\,,
\end{equation}
with the solution
\begin{equation}
    \rho_t^{(gg)} = 1 - \rho_0^{(ee)} e^{-\Gamma t} \,,
\end{equation}
meaning an exponential decay of the excited state population with rate $\Gamma$, as it can be observed in \cref{fig:spontemiss_SSE_dynamics}.
Note that, as the normalization is ensured \textit{on average} in the stochastic formulation, in this case we can clearly observe that the single trajectories are not individually normalized. 
\vspace{1ex}

At this point, some remarks on the choice of the noise source are in order. While we start without making any assumption on the noise term in \cref{eq:Ito_SSE_GENERAL}, the developments of this subsection explicitly rely on the white noise assumption. Indeed, the Lindblad form is obtained only when using white noise, a Gaussian $\delta$-correlated distribution, i.e., that retains no memory of its past evolution.
This corresponds to a full Markovian approximation, the same that is required in the derivation of the Lindblad form from any other route \cite{Lindblad1976MathematicalSemigroups,Gorini1976CompletelySystems,petruccione2002}. 
In particular, within the white noise assumption, no constraints arise for the operator $B$ to satisfy the normalization condition \cref{eq:normal_A,eq:Aoperatorwithcorrection}. 
We stress the fact that this is a crucial point allowing the connection of a linear SDE to all possible master equations of Lindblad form. 

When dealing with different noise-inducing terms, different forms of master equations arise, introducing non-Markovian terms into the dissipator. Indeed, abandoning the white noise assumption in the SSE changes significantly its properties. For example, to preserve the linear formulation presented above, additional constraints on the operators must be introduced. Most notably, to maintain the linearity of the SSE, the operator $B$ must be anti-Hermitian, leading to a stochastic Hamiltonian interpretation of the SSE. A clarification of this concept is discussed in the following section, while we refer the reader to ref \cite{DeChecchi2025DynamicsEquations} for an in-depth analysis of the effects of colored source of noise. Alternatively, different routes must be considered, such as non-linear formulations \cite{Barchielli2009Quantum782,Barchielli2010StochasticNoise}.

\subsection{The stochastic Hamiltonian as an instance of the stochastic \schr equation}

Since every Lindblad form can be obtained from an SSE using white noise by an appropriate choice of the operator $B$, it is natural to wonder whether the stochastic Hamiltonian can be regarded as a specific case, a subclass, of this stochastic approach.
Comparing the stochastic Hamiltonian formulation of \cref{eq:stoch_schr_eq_gen} with the general form of \cref{eq:Ito_SSE_GENERAL}, one may relate the two noise coupling operators as $B=iR$. Notice that the stochastic Hamiltonian needs the additional constraint 
$R\daggah=R$, to preserve the hermiticity of the effective Hamiltonian.
However, this form prevents us from simply deriving the exponential solution as done previously in \Cref{sec:stratonovich}.
Indeed, the incorrect exponentiation would lead to real negative contributions, causing the loss of wavefunction norm at each time step.
This is due to the interpretation of the stochastic differential, which does not allow for ordinary calculus rules. 
Therefore, to be able to obtain the correct unitary map \cref{eq:unitary_propagator}, we need to change the interpretation of the stochastic differential from the \ito form ($\diff{W_t}$)
to the Stratonovich one ($\circ\diff{W_t}$), as done in \Cref{sec:stratonovich}.

Here, we show how to move from Stratonovich to \ito for the stochastic Hamiltonian class,
hence recovering the Lindblad master equation, and
showing the equivalence of the two methods.

Starting from the stochastic Hamiltonian definition in the Stratonovich formalism in \cref{eq:Stratonovich_SSE}, using the definition of the Q-multiplication in \cref{eq:stratonovich_product}, we can write for the stochastic component in the wavefunction differential
\begin{equation}
\label{eq:intermediate1_stran-ito}
    -i\gamma R\psi_t\circ\diff W_t = -i\gamma R\psi_t\diff{W_t} -\frac{i\gamma}{2}R\diff\psi_t\diff{W_t},
\end{equation}
where we insert back the definition \cref{eq:Stratonovich_SSE} of the differential $\diff{\psi_t}$ to obtain 
\begin{equation}
\label{eq:intermediate2_stran-ito}
\begin{split}
    -i\gamma R\psi_t\circ\diff W_t = &{-i}\gamma R\psi_t\diff{W_t} \\[.5ex]
    &-\frac{i\gamma}{2}R (-iH\psi_t\diff{t} - i \gamma R\psi_t\circ\diff{W_t}) \diff{W_t} \,.
\end{split}
\end{equation}
To avoid the inconsistency of dealing with mixed \ito and Stratonovich differential terms, we need to apply again \cref{eq:intermediate1_stran-ito} to the Stratonovich term in the parentheses above obtaining, for the r.h.s. of \cref{eq:intermediate2_stran-ito},
\begin{equation}
\label{eq:intermediate3_stran-ito}
\begin{split}
    &-i\gamma R\psi_t\diff{W_t}
    -\frac{i\gamma}{2}R 
    \Bigg\{ 
    -iH\psi_t\diff{t} - i \gamma R\ \times \\
    &\times\left[
        \psi_t\diff{W_t} +\frac{1}{2} \left(-iH\psi\diff{t} -i\gamma R \psi\circ\diff{W_t}\right)
    \right]
    \Bigg\}
    \diff{W_t},
\end{split}
\end{equation}
where the terms of order $\mathcal{O}(t^{3/2})$ and higher \footnote{$\mathcal{O}(\diff{t}\diff{W})=\mathcal{O}(t^{3/2})$, $\mathcal{O}(\diff{t}\diff{t})=\mathcal{O}(t^{2})$, and since $\mathcal{O}(\diff{t}\diff{W}\circ\diff{W})>\mathcal{O}(t^{3})$, any terms arising from a further expansion are evidently of order higher than $\mathcal{O}(t^{3/2})$.}
can be neglected. 
Then, the only terms that survive are the ones in $\diff{W}$ and $(\diff{W})^2$: 
\begin{equation}
\label{eq:intermediate_last_stran-ito}
    -i\gamma R\psi_t\circ\diff W_t = -i\gamma R\psi_t\diff{W_t} -\frac{\gamma^2}{2}R^2\psi_t\diff{t}.
\end{equation}
Notice that the last term, proportional to $\gamma^2$, corresponds indeed to the normalization term required in the SSE approach using the \ito formalism. 
In the Stratonovich formalism for a stochastic Hamiltonian, the normalization is implied \textit{a priori}, and does not need to be enforced, as it comes with the formalism, and the normalization term arises when rewriting into the \ito formalism.

The one additional constraint to be imposed for the normalized SSE in \ito formalism, \cref{eq:Ito_SSE_whitenorm}, to be equivalent to the Stratonovich stochastic Hamiltonian dynamics, \cref{eq:Stratonovich_SSE}, 
is that the operator $R$ must be Hermitian, \cref{eq:hermitian_R}, 
together with the identification $B=iR$ anticipated at the beginning of this section.
With this constraint, substituting \cref{eq:intermediate_last_stran-ito} into \cref{eq:Stratonovich_SSE}, we obtain the norm-preserving SSE, \cref{eq:Ito_SSE_whitenorm}, and then by usual \ito's calculus we obtain again QMEs in Lindblad form.
Notice that the additional condition on $B$, which is required to be anti-Hermitian in this case, defines a specific subclass of Lindblad dynamics, those unraveled by stochastic unitaries. Notable examples are pure decoherence dynamics and population transfer in the high temperature limit \cite{Gu2019WhenNoise}. 

We conclude that, within this specific subclass of the SSE, one can use one formalism or the other to their convenience.
Using Stratonovich for deriving a stochastic evolution in the form of a closed standard Hamiltonian dynamics and admitting fluctuations in the Hamiltonian, leads to a structure more immediate to understand, and that allows for direct quantum computing implementation.
The construction from \ito formalism, leading to the SSE, is otherwise used for the analytical derivation of quantum master equations, and allows an efficient numerical implementation as further explained in the following.
As long as the constraints presented for the equivalence of the two formulations hold, the transformation between them is well-defined, and one can use the description most suited to their needs.

\subsection{Advantages of the \ito framework}

The first clear advantage of the SSE approach and the \ito interpretation is that it allows the formal derivation of the associated QMEs.
This is true independently of the form of the operators.
Indeed, we can obtain a variety of dynamics much wider than the sole stochastic Hamiltonian class. 
Restricting ourselves to this smaller class, we still have flexibility with the choice of the noise-inducing differential $\diff{X_t}$,
and we can generalize to non-linear SSE, and to various measurement interpretations \cite{Barchielli2009Quantum782}.

As of today, with quantum computations not yet living up to their full potential, the other important advantage of the SSE framework is that it allows efficient implementation schemes on classical computing architectures, avoiding the necessity to implement matrix exponentiation at each timestep of propagation and relying on efficient linear algebra schemes. 
These schemes are based on finite-difference integration, such as the
Euler–Maruyama method \cite{Bally1995TheCalculus,Bally1996TheFunction},
the extension to SDE of the Euler finite difference methods,
\begin{equation}
    \psi_{n+1} = \psi_{n} - \left(iH +\frac{1}{2}\gamma^2 B\daggah B \right)\psi_n \Delta t 
    +\gamma B \psi_t \Delta W_n.
\end{equation}

Other integration schemes are the Milstein method, which adds a correction to the strong order of convergence \cite{Milstein1995NumericalEquations},
$\mathcal{O}(\Delta t)$ instead of $\mathcal{O}(\sqrt{\Delta t})$ but maintaining the weak order of convergence,
and Runge-Kutta methods, less used in the stochastic setting as their complexity rapidly increases with the order of the method \cite{Platen1992NumericalEquations,Mora2004NumericalSemigroups,Platen2010NumericalFinance,Roler2009SecondEquations,Roler2010Runge-KuttaEquations}.

We should point out that quantum computing implementation of the dynamics obtained in the SSE method is still possible, but with some caveats. The implementation through stochastic unitaries as described in \cref{sec:unitary_quantumalg} is possible as long as the system can be described as a stochastic Hamiltonian, and therefore recast in its Stratonovich equivalent. Otherwise, different schemes must be used and relevant examples can be found in refs.~\cite{Hu2020ADevices,Sweke2015UniversalSystems,Sweke2016DigitalDynamics,Schlimgen2022QuantumOperators, Miessen2022QuantumDynamics,MacDonell2023PredictingSimulation}. 

\section{Stochastic Hamiltonians with continuous stochastic processes}\label{sec:rode}

In the sections above, we used two definitions of stochastic differentials, \ito and Stratonovich, to introduce the effect of noise on the quantum dynamics.
On the other hand, the fluctuation in the Hamiltonian in \cref{eq:eff_stoch_hamiltonian}
can also be a stochastic \textit{process}, namely a continuous function that solves its own SDE.
In this case, $Z_t$ in \cref{eq:eff_stoch_hamiltonian} is a well-defined continuous function and not a distribution, therefore \cref{eq:stoch_schr_eq_gen} 
can be treated as an ordinary differential equation (ODE) with random and fluctuating coefficients (RODE), without the need to resort to the stochastic differential equation frameworks.

This is equivalent to setting to zero the stochastic differential terms in \cref{eq:Ito_SSE_GENERAL,eq:Stratonovich_SSE} and letting random fluctuations appear in the operator $A$, which now becomes random and time-dependent. In this way, we obtain precisely a \schr equation with the stochastic Hamiltonian in \cref{eq:stoch_schr_eq_gen}.
The evolution of the system wavefunction is then well defined, continuous, and at least one time differentiable. 
This special case allows for the use of random processes in the effective Hamiltonian, such as the Ornstein-Uhlenbeck process (OU)  \cite{Uhlenbeck1930OnMotion,Gillespie1996ExactIntegral}, and not their noisy increments \cite{Gallina2024FromDynamics,DeChecchi2025DynamicsEquations}.
The OU process is a Gaussian stochastic process with zero mean characterized by a two-time correlation function decaying exponentially,
see \cref{app:whitenoise}.
This process has been used to study memory effects in excitation transport in chromophore aggregates \cite{Fujita2012Memory-AssistedBacteria,Dijkstra2015CoherentVibrations,Gallina2024FromDynamics},
and to describe random modulation of the spin Hamiltonian in magnetic resonant experiment \cite{Kubo1969AShape}. We will return to this latter application in the illustrative example below.

Since the problem of the dynamics is well posed, the numerical implementation inherits the advantages of both methods discussed above: it can be discretized in time and integrated \textit{via} finite-difference schemes to be optimized in classical computers. 
It also admits the unitary propagator, as long as the constraints of the stochastic Hamiltonian hold, which allows quantum computing implementation \textit{via} unitary gates.

\begin{figure}
    \centering
    \includegraphics[width=\linewidth]{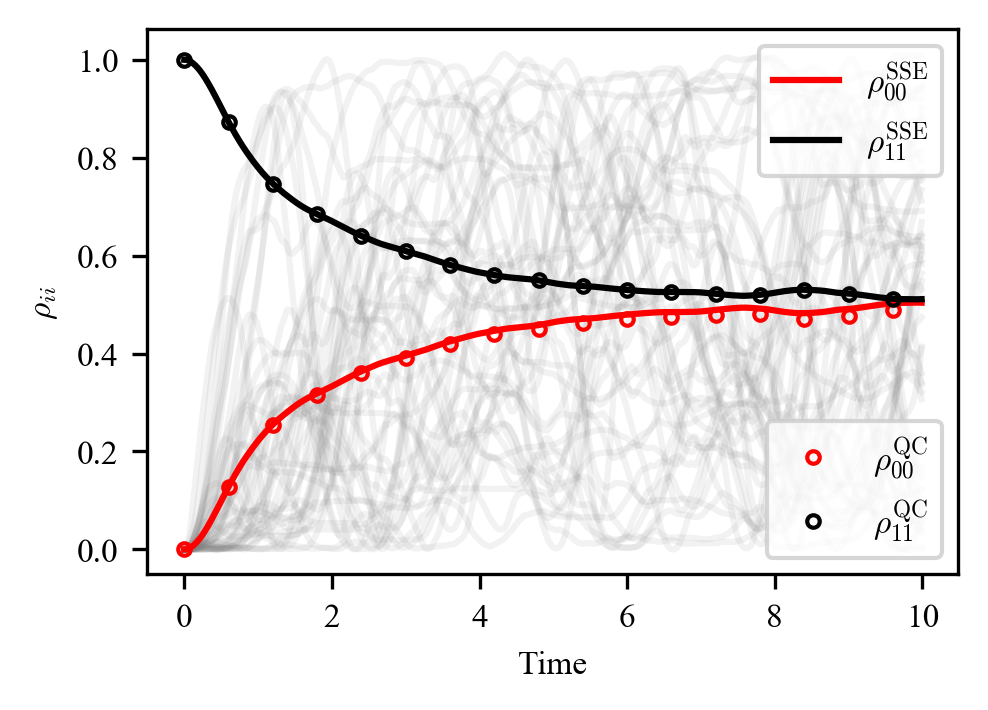}
    \caption{Relaxation of the population of a $\frac{1}{2}$-spin populations along the $z$-axis, obtained by averaging over $10^3$ trajectories computed in the RODE stochastic Hamiltonian approach (an ensemble of $10^2$ trajectories is shown in gray transparency lines). The average for the two populations is shown in solid red and black lines.
    The empty dots, with the same color code as the averages, show the results of a quantum algorithm realization, for $10^3$ trajectories, each point sampled by 1024 ``shots''.
    }
    \label{fig:SpinTumbl}
\end{figure}

\vspace{1ex}
\textbf{Example 4 (spin-$\frac{1}{2}$ molecular tumbling)} 
Continuing with the use of a two-level system and prototypical examples, let us consider the description of 
spin relaxation channels, relevant in magnetic resonance experiments such as NMR (nuclear magnetic resonance) and EPR (electron paramagnetic resonance).
Take a single $\frac{1}{2}$-spin system in a static magnetic field $\Phi_0$ along the $z$-axis direction.
This system, in the absence of molecular motions, 
is described by the Hamiltonian
\begin{equation}
    H_0 = \frac{\omega_0}{2}\sigma_z
\end{equation}
where $\omega_0\propto\Phi_0$ is the Larmor frequency of the system.
In liquid solutions, the molecule undergoes translational and rotational diffusion. These motions imply random modulations of the spin Hamiltonian that can be modeled by a real stochastic process. To account for a finite correlation time, the  Ornstein-Uhlenbeck model mentioned above is often used   \cite{Kubo1970BrownianSpins,Abragam1983PrinciplesPhysics,Field2013DynamicalRelaxation}, which we denote here as $X_t$.
The final effective Hamiltonian takes then the form
\begin{equation}
    H_t^\mathrm{eff} = \frac{\omega_0}{2}\sigma_z + \mathbf{X}_t \cdot \boldsymbol\sigma \,
\end{equation}
where $\boldsymbol{\sigma}=(\sigma_x,\sigma_y,\sigma_z)$ is a vector of Pauli matrices for the three axis and $\mathbf{X}_t=(X^{(x)}_t,X^{(y)}_t,X^{(z)}_t)$ is a vector of OU processes representing the fluctuation of the local field experienced by the spin due to the rotational diffusion.  
The system evolution is solved as an RODE,
propagated with finite-difference numerical methods, either by linearization with Euler method, Runge-Kutta methods, or, with the analytical exponential, \cref{eq:unitary_propagator},
by time-discretization in either classical or quantum computers, \cref{eq:propagator_discretization}.
The solution for the average spin populations is shown in \cref{fig:SpinTumbl}, together with the emulation of a quantum circuit using a sequence of unitaries as presented in \cref{sec:unitary_quantumalg}, adapted using the OU fluctuations on all three Pauli rotation gates $R_X,\, R_Y$ and $R_Z$.

\section{Stochastic Hamiltonian in  Liouville Equations}\label{sec:sle}

The previous sections rely on and exploit the use of stochastic state vectors.
This allows, on the one hand, to use non-Hermitian operators in the case of \ito SDE formalism and obtain any QME in Lindblad form, and
on the other hand, using Stratonovich interpretation, to obtain unitary evolutions of pure states which can be implemented on digital quantum computing architectures.
In a further framework, one can directly work with the system density matrix and insert stochastic fluctuations into its dynamics.
This can be done for various purposes, allowing for both different descriptions and different numerical implementations on either classical or quantum computers.

Several works have explored this approach, producing interesting results 
\cite{Budini2000Non-MarkovianVector,Budini2001QuantumFields,Kiely2021ExactConnections,Chenu2017QuantumNoise}.
However, the derivations are often presented concisely, with some assumptions and intermediate steps left implicit for the reader. This can lead to differing results, due to tacit approximations. 
In what follows, we aim to complement these contributions by providing a detailed derivation of different master equations, starting from the Liouville–von Neumann form with a stochastic component, and by clarifying the role of assumptions and approximations.
First, we obtain non-Markovian forms of the master equations and then, through different approximations, we recover the results of Ref. \cite{DeChecchi2025DynamicsEquations}, which link these stochastic approaches to the Redfield master equation. 
Finally, we show how further approximations lead to Lindblad forms.
A graphical outline of our presentation of the SLE approach, highlighting its main directions, is shown in \cref{fig:SLE_scheme}.

\begin{figure}
    \centering
        \begin{tikzpicture}[node distance=2cm, >=Stealth]
            \tikzstyle{block} = [draw, rounded corners, minimum width=2.2cm, minimum height=1.2cm, align=center]
            \tikzstyle{circleblock} = [draw, circle, minimum width=2.2cm, align=center]
            \node (commutator) {$\left[\hat H^{\text{fluct}}_t,\, \rho_t^{\text{traj}}\right]$};
            \node[below=1cm of commutator] (dummydown) {};
            \node[right=1cm of commutator] (exact) {$\frac{\mathrm{d}\rho_t}{\mathrm{d}t}$};
            \node[right=2cm of exact] (rho) {$\rho_t$};
            \node[below=1cm of exact] (approx) {$\frac{\mathrm{d}\rho^\mathrm{model}_t}{\mathrm{d}t}$}; 
            \node[right=2cm of approx] (rhotilde) {$\rho^\mathrm{model}_t$};
            \draw[->,midway,above] (commutator) to node {$\mathbb{E}$} (exact);
            \draw[->,above] (exact) to node {\text{numerical}} (rho);
            \draw[->,above] (approx) to node {\text{analytical}} (rhotilde);
            \draw[->,midway,sloped,above] (exact) to node {$\approx$} (approx);
        \end{tikzpicture}
        \caption{A schematic map of the stochastic Liouville equation (SLE) framework, where averaging the commutator dynamics yields the exact evolution of the density matrix (upper branch), which can be solved numerically, while approximate models provide analytical solutions (lower branch).}
    \label{fig:SLE_scheme}
\end{figure}

For the stochastic Hamiltonian of \cref{eq:eff_stoch_hamiltonian}, we can write the stochastic Liouville equation (SLE) as
\begin{equation}
\label{eq:SLE}
    \frac{\diff}{\diff{t}}\rho^\mathrm{trj}_t = -i [H^\mathrm{eff}_t, \rho^\mathrm{trj}_t]
    = -i [H, \rho^\mathrm{trj}_t] -i [Z_t R, \rho^\mathrm{trj}_t],
\end{equation}
of which we can take the average to obtain the mean dynamics given by
\begin{equation}
\label{eq:mean_SLE}
    \frac{\diff}{\diff{t}}{\rho}_t
    = -i [H, {\rho}_t] -i [R,\mathbb{E}(Z_t\rho^\mathrm{trj}_t)] 
\end{equation}
where, thanks to $Z_t$ being a real-valued one-dimensional process and $R$ a deterministic operator, we can bring the average on the product $Z_t\rho^\mathrm{trj}_t$ inside the commutator.
Since this term depends on the explicit average of the trajectory-wise product over the entire sample space, 
\cref{eq:average},
$\mathbb{E}(Z_t\rho^\mathrm{trj}_t)$ is an open term
that must be treated carefully.
Due to its presence, \cref{eq:mean_SLE} is not a closed-form master equation in terms of $\rho_t$ only, but explicitly depends on $\rho^\mathrm{trj}_t$ and $Z_t$.

As the stochastic density matrix is a functional $\rho^\mathrm{trj}[Z_t]$ of the random variable $Z_t$, under the important assumption that $Z_t$ is Gaussian, we can apply Novikov's theorem
\cite{Novikov1964FunctionalsTheory, Klyatskin1973StatisticalSystems}
which, for the quantities above, gives 
\begin{equation}
\label{eq:Novikov_thm}
    \mathbb{E}(Z_t\rho^\mathrm{trj}_t) 
    = \mathbb{E}(Z_t)\mathbb{E}(\rho^\mathrm{trj}_t) 
    + \int_{t_0}^t \mathbb{E}(Z_t Z_s) 
    \mathbb{E}\left(\frac{\delta\rho^\mathrm{trj}_t[Z]}{\delta Z_s}\right)\diff s  
\end{equation}
where, for zero-mean processes, the first term on the r.h.s. vanishes 
and the first term in the integral is the covariance of the process, $C_Z(t,s) = \mathbb{E}(Z_t Z_s)$.
It now remains to compute the functional derivative in the integrand.
First, we write the formal solution of \cref{eq:SLE} as
\begin{equation}
\label{eq:intermediate-1_Novikov}
    \rho^\mathrm{trj}_t = \rho^\mathrm{trj}_{t_0} -i\int_{t_0}^t [H+Z_r R, \rho^\mathrm{trj}_r] \diff{r}
\end{equation}
and then we take the functional derivative to obtain
\begin{align}
\label{eq:intermediate-2_Novikov}
\begin{split}
    \frac{\delta\rho^\mathrm{trj}_t[Z]}{\delta Z_s} = -i\int_{t_0}^t 
    \Bigg(
    &\left[    H+Z_r R, \frac{\delta\rho^\mathrm{trj}_r[Z]}{\delta Z_s} \Theta(r-s) \right]\\[.5ex]
    &+ \delta(s-r) \left[ R, \rho^\mathrm{trj}_r \right] \Bigg)\diff{r},
\end{split}
\end{align}
where $\delta(s-r)$ is the Dirac delta function and $\Theta(r-s)$ is the Heaviside function for the shifted time. Due to the definition
\begin{equation}
    \Theta(r-s)=\left\{
    \begin{array}{ll}
        0 & r<s \\
        1 & r\geq s
    \end{array}
    \right.
\end{equation}
the functional derivative of $\rho^\mathrm{trj}_r$ does not depend on its future, and we can change the inferior limit of the integral and arrive at
\begin{equation}
    \label{eq:intermediate-2.2_Novikov}
    \frac{\delta\rho^\mathrm{trj}_t[Z]}{\delta Z_s}
    = -i  \left[ R, \rho^\mathrm{trj}_s \right] -i\int_{s}^t 
    \left[    H+Z_r R, \frac{\delta\rho^\mathrm{trj}_r[Z]}{\delta Z_s}\right]\diff{r} .
\end{equation}

It seems that we have replaced one problem with another.
Yet, taking the time derivative, we can write
\begin{equation}
\label{eq:intermediate-3_Novikov}
    \frac{\diff}{\diff{t}} \frac{\delta\rho^\mathrm{trj}[Z]}{\delta Z_s} = -i\left[H+Z_t R, \frac{\delta\rho^\mathrm{trj}_t[Z]}{\delta Z_s}\right],
\end{equation}
where we can recognize the form of a Liouville differential equation, $\dot{A}=-i[H^\mathrm{eff}_t,A]$, where $A$ is the functional derivative of the trajectory density matrix.
From this recasting, we thus obtain a known form of the solution, $A_t=U{A_0}U\daggah$, with the unitary time-evolution operator defined by
\begin{equation}
    \mathcal{U}_{t,t_0} = \timeorder \exp\left\{ -i \int_{t_0}^t (H+Z_r R) \diff{r} \right\}.
\end{equation}
The initial condition is obtained by evaluating \cref{eq:intermediate-2.2_Novikov} at the initial time (which is $s$) and corresponds to the first term in \cref{eq:intermediate-2.2_Novikov}, as the integral vanishes when computed at a single time.
Finally, we can write
\begin{equation}
\label{eq:novikov_final_differentialsolution}
    \frac{\delta\rho^\mathrm{trj}[Z]}{\delta Z_s} 
    = \mathcal{U}_{t,s} \big(-i \left[ R, \rho^\mathrm{trj}_s \right] \big)\mathcal{U}_{t,s}\daggah
    = - i\left[ \mathcal{U}_{t,s}R\mathcal{U}_{t,s}\daggah, \rho^\mathrm{trj}_t \right] 
\end{equation}
where the propagator is distributed in the commutator and acts on the density matrix, evolving it from the initial time $s$ to the final time $t$.
This shows that the propagation 
can be dealt with as a time-local evolution with respect to the trajectory density matrix, a time-convolutionless evolution \cite{Budini2001QuantumFields}.
We can now take the average and combine with \cref{eq:Novikov_thm,eq:mean_SLE}, yielding
\begin{equation}
\label{eq:mean_SLE_nM}
\begin{aligned}
    \frac{\diff}{\diff{t}}{\rho}_t
    = &-i [H, {\rho}_t] \\
    &- \int_{0}^t C_Z(t,s) \left[R,   \mathbb{E}\left( \left[ \mathcal{U}_{t,s}R\mathcal{U}_{t,s}\daggah, \rho^\mathrm{trj}_t \right] 
    \right) \right] \diff{s}
    \end{aligned}
\end{equation}
as the evolution equation for the mean density matrix.
It is still in an open integro-differential and non-Markovian form for a generic noise or process $(Z_t)$, and solvable only numerically through a swarm of trajectories and their average.
One exception is the case of white noise, which we will treat later, for which there are further simplifications.
Then, for different sources of randomness of the Hamiltonian, such as colored processes and noises, one has to rely either on the numerical implementation, or on the use of perturbative approximation schemes.

For example, a closed deterministic master equation can be obtained
using a perturbative approach, where the propagator is simplified according to 
\begin{equation}
\label{eq:approx_propagator}
\begin{split}
    \mathcal{U}_{t,t_0} 
    &= \timeorder \exp\left\{ -i \int_{t_0}^t (H+Z_r R) \diff{r} \right\} \\[1ex]
    &\approx \exp\left\{ -i \int_{t_0}^t H \diff{r} \right\} = \mathcal{U}_{t,t_0}^\mathrm{H}.
\end{split}
\end{equation}
Then, we can write $R_{t,s} = \mathcal{U}_{t,s}^\mathrm{H}R\mathcal{U}_{t,s}^\mathrm{H\,\dagger}$, now a deterministic operator independent on the stochastic process, therefore not affected by the average operator.
This allows us, by linearity of the average, to insert the averaging operator $\mathbb{E}[\cdot]$ in the commutator in \cref{eq:mean_SLE_nM} to act solely on the ensemble of stochastic density matrix trajectories $\rho^\mathrm{trj}$ and thereby close the equation in terms of the average trajectory, as $\mathbb{E}[\rho_t^\mathrm{trj}]=\rho_t$.

Evaluating the action of the approximated two-times propagator on the noise operator, one obtains a QME with \textit{time-dependent coefficients} identical to what is obtained while applying a Redfield approach
\cite{DeChecchi2025DynamicsEquations}:
\begin{equation}
    \label{eq:redfield}
    \frac{\diff}{\diff{t}}{\rho}_t
    \approx -i [H, {\rho}_t] - \int_{0}^t C_Z(t,s) \left[R,   \left[ R_{t,s}, \rho_t \right]  \right] \diff{s} .
\end{equation}
The last step would be to apply the complete Markov approximation by extending the integration upper bound to infinity, eventually recovering the well-known Redfield master equation.

In the particular case of white noise, one can straightforwardly obtain a closed-form master equation, as the correlation function is now a delta function.
In fact, the integral term greatly simplifies: the integral is evaluated at time $t$, where the propagator effects vanish, and the average then acts only on the stochastic density matrix, giving the mean $\rho_t$, which undergoes a local-in-time Markovian evolution.  
As expected, this leads to the Lindblad form
\begin{equation}
    \frac{\diff}{\diff{t}}{\rho}_t
    = -i [H, {\rho}_t] - \left[R, \left[ R, {\rho}_t \right] \right].
\end{equation}

To summarize, within this theoretical framework and under some constraints
(such as the Gaussianity of the real-valued stochastic term and the noise operator being deterministic), we can obtain the exact evaluation of the cross-correlation between the system dynamics and the process modeling the environment, see \cref{eq:Novikov_thm}, and most notably its expression in the mean system dynamics, \cref{eq:mean_SLE_nM}. 
Although readily computable only for white noise sources, it allows for closure models under some approximations,
e.g.\ \cref{eq:approx_propagator} leading to \cref{eq:redfield}, reconciling the results previously obtained in 
\cite{DeChecchi2025DynamicsEquations}.
We further note that the QMEs in open forms, \cref{eq:mean_SLE_nM}, obtained by the stochastic approach are the average of pure states.
Therefore, they ensure positivity of the mean density matrix by the same arguments introduced in \cref{sec:modeling_SH}. 
If we introduce approximations, such as the perturbative approach through \cref{eq:approx_propagator}, positivity is not ensured anymore.
In fact, in that case, we obtain a Redfield form, \cref{eq:redfield}, in which the loss of positivity is well-known.

\subsection{Advantages of implementing density matrix dynamics}

Although simple systems with only a stochastic Hamiltonian could be propagated more efficiently through wavefunction methods, the SLE methods allow for a wider range of system dynamics.
They allow us to consider the effects of stochasticity on general mixed states and implement systems undergoing different non-unitary irreversible dynamics.
In a general form as follows,
\begin{equation}
\label{eq:additional_diss_sle}
    \frac{\diff}{\diff{t}}\rho^\mathrm{trj}_t = -i [H^\mathrm{eff}_t, \rho^\mathrm{trj}_t] + \mathcal{D}[\rho^\mathrm{trj}_t] + \mathcal{F}[\rho^\mathrm{trj}_t] \,,
\end{equation} 
we can consider $\mathcal{D}[\tilde{\rho_t}]$ a different dissipator acting on the trajectory density matrix, in Lindblad or Redfield form, or any other, and $\mathcal{F}[\rho^\mathrm{trj}_t]$ to be any filtering, measurement schemes, and any other component we want to study in a particular and complex dynamics. 
It also allows us to perform these general dynamics on systems and then trace out part of it, allowing for three different layers of interactions: a purely non-Markovian one (the component that is traced out), irreversible dissipation and measurements on the trajectories, stochasticity on the Hamiltonian of the single realizations leading to different dissipators in the mean dynamics.

These dynamics, either the simple stochastic Hamiltonian or the more complex one, can be implemented both classically, mainly for small systems, and on quantum computers, through  
dilations of the Hilbert space and decompositions into Kraus OSR maps 
\cite{Kraus1971GeneralTheory,Kraus1983StatesTheory, Hu2020ADevices,Sweke2015UniversalSystems,Sweke2016DigitalDynamics,Schlimgen2022QuantumOperators}.

\section{Conclusions}\label{sec:conclusions}

In this work, we provide a unified and detailed description of different modeling techniques for open quantum systems,
including, but not limited to, the rich and flexible framework of stochastic Hamiltonians, a deliberately phenomenological approach introduced as direct modeling tools to capture environmental effects without reference to an explicit microscopic derivation. In this spirit,  these approaches are not derived from a unitary total system–bath model but are rather assumed as an effective description, chosen to reproduce the relevant dynamical and spectroscopic features. From this foundation, we trace how different formulations—ranging from stochastic Hamiltonians to stochastic Schrödinger and Liouville equations—can be related and systematically interpreted.
We highlight the implementation advantages and theoretical constraints of these frameworks for quantum and classical computation. 
Two interpretations of stochastic calculus, \ito and Stratonovich approaches, are relevant to different aspects of the dynamics of open quantum systems.

In the Stratonovich formalism, the rules of ordinary calculus are maintained.
This allows us to write the propagator of the dynamics in analogy with the closed-system \schr equation.
Being a unitary operator, it yields unitary evolutions, ensuring norm preservation trajectory-wise.
This makes this method well-suited for the implementation in quantum computers.

On the other hand, within the \ito interpretation, we can derive the associated quantum master equations, which govern the observed mean dynamics of the open system.
Beyond the stochastic Hamiltonian, we have shown, with the example of the derivation of the generic QME in Lindblad form, that the stochastic \schr equation method obtained in this formalism encompasses a broader class of open system dynamics, and stochastic Hamiltonians emerge as one specific subclass. 
While most applications consider white noise, this approach can be extended to the use of colored and unconventional noises as stochastic driving, as shown, for example, in Refs.\ \cite{Barchielli2010StochasticNoise,Cialdi2019ExperimentalDynamics,DeChecchi2025DynamicsEquations,DeKeijzer2025QubitNoise}.
Although this method does not ensure norm-preservation for each trajectory, but only on average, it allows for efficient numerical integration schemes on classical architectures.

Additionally, we discussed the use of RODEs as a special case where the Hamiltonian includes continuous stochastic processes. When applicable, this stochastic Hamiltonian approach features both unitary evolution at the trajectory level, useful for quantum computing implementation, and a form amenable to several efficient integration schemes in classical architectures. 

Lastly, we discussed the stochastic Liouville equation,
as it provides an alternative by working directly with density matrices and enabling different ways to include additional dissipative channels and measurement processes.

We show how to formally move between different interpretations of the stochastic terms and the related methods.
These methods altogether offer a versatile toolkit for the modeling and simulation of open quantum systems.
The choice of the method needs to be guided both by the desired physical insight and by the computational platforms or analytical tools one wishes to employ.

\section*{Acknowledgments}

The authors thank M. Bruschi for fruitful discussions. 
P.D.C., G.G.G.\ and B.F.\ acknowledge funding from the European Union - NextGenerationEU, within the National Center for HPC, Big Data and Quantum Computing (Project no. CN00000013, CN1 Spoke 10: “Quantum Computing”).
B.F. and F.G.\ acknowledge the financial support by the Department of Chemical Sciences (DiSC) and the University of Padova with Project QA-CHEM (P-DiSC No. 04BIRD2021-UNIPD).

\appendix

\section{White noise, Brownian motion and Ornstein-Uhlenbeck process}\label{app:whitenoise}
White noise $\xi_t$ is an idealization of interactions with environments with correlation times much shorter than the characteristic time of the system, which can therefore be considered memoryless.
It can be heuristically defined as the formal derivative of the Wiener process $W_t$ (Brownian motion) 
$\xi_t=\gamma\diff{W}_t/\diff{t}$.
This is a distribution, as $W_t$ is not differentiable. 
This noise has the following properties:
\begin{enumerate}[label=(\roman*)]
    \item its mean is null, $\mathbb{E}[\xi_t]=0$,
        and
    \item it is $\delta$-correlated in time, $\mathbb{E}[\xi_t\xi_s]=\gamma^2\delta(t-s)$,
\end{enumerate}
where $\gamma$ is the intensity of the noise and $\delta(t-s)$ is the Dirac delta at different times. 
Its properties, and those of the differential $\diff{W_t}$, follow from the definition of the Wiener process $W_t$,
a real-valued Gaussian stochastic process with independent and stationary increments, with zero mean and variance proportional to the time $t$. Equivalently, $W_t$ is bestowed with the following properties:
\begin{enumerate}[label=(\roman*)]
    \item $W_0=0$ a.s.,
    \item $W_t-W_s=\mathcal{N}(0,t-s) \;\forall{t}\geq{s}\geq0$, and
    \item $(W_t-W_s)$ is independent of $(W_s-W_r)$ $\forall{t}\geq{s}\geq{r}\geq0$.
\end{enumerate}
We can write for its differential the following useful properties,
\begin{enumerate}[label=(\roman*)]
    \item $\diff{W_t}\sim\mathcal{N}(0,\diff{t})$,
    \item $\mathbb{E}[\diff{W_t}]=0$, and
    \item $\mathbb{E}[\diff{W_t}^2]=\diff{t}$,
\end{enumerate}
that are used throughout the main body of the paper.
The definition of \textit{white} noise comes from the spectral density of the process, i.e., the Fourier transform of its autocorrelation function, which is constant, meaning that all the frequencies of the environment are contained in and contribute equally to the correlation function.

A generalization of the white noise is the Ornstein-Uhlenbeck process $X_t$, characterized by the following SDE,
\begin{equation}
    \diff{X_t} = -\theta X_t \diff{t} + \gamma\diff{W_t}\,,
    \label{eq:OU_differential}
\end{equation}
where $\theta$ is the inverse of the correlation time of the process and $\gamma$ is the intensity of the underlying white noise fluctuation.
Therefore, in the case of $\theta=0$, we recover a white noise of intensity $\gamma$.
The integral form of the process, used as the fluctuation, is 
\begin{equation}
    \label{eq:OU_integralform}
    X_t = X_0 e^{-\theta t} + \gamma\int_0^te^{-\theta (t-s)}\diff W_s\,.
\end{equation}
This process is characterized by a two-time correlation function decaying exponentially,
\begin{equation}
    c_\mathrm{OU}(t,t_0)
    = \frac{\gamma^2}{2\theta} e^{-{|t-t_0|}{\theta}}\,,
\end{equation}
related to a Lorentzian spectral density in the form
\begin{equation}
    \label{eq:SD_OU-Lorentzian}
    J_\mathrm{OU}(\omega) = \int_{-\infty}^{+\infty} c_\mathrm{OU}(t,t_0) e^{i\omega t} \diff t = \frac{\gamma^2}{\omega^2+\theta^2}\,.
\end{equation}

\section{Solution of the Haken-Strobl model}\label{app:HS_Lindblad}
The Lindblad master equation representing the exact solution of the Haken-Strobl model discussed in the example of \cref{sec:modeling_SH} is
\begin{equation}
    \difft\rho_t = -i[H,\rho_t] + 
    \sum_{i=0}^1
    \Gamma_i\left( \ketbra{i}{i}\rho_t \ketbra{i}{i} -\frac{1}{2} \left\{\ketbra{i}{i},\rho_t\right\} \right)
\end{equation}
where $[\cdot,\cdot]$ is the commutator, $\{\cdot,\cdot\}$ the anticommutator, and $\Gamma_i = \mathbb{E}[\delta_i^2(t)]$ the dephasing rate associated with site $i$.
We recall that $\delta_i(t)$ are intended as white noises.

This serves as an exact reference for the convergence of the density matrix obtained by averaging over a finite sample space.
While the Hamiltonian term of the equation drives the coherent oscillations between the sites, the effect of the dephasing term is to destroy such coherences.
As a result, the off-diagonal terms of the density matrix decay exponentially
\begin{equation}
    \difft\rho^{(01)}_t = -i2\Omega \left(\rho^{(11)}_t - \rho^{(00)}_t\right) - \frac{\Gamma_0 + \Gamma_1}{2} \rho^{(01)}_t \,,
\end{equation}
where the superscript $(ij)$ indicates the density matrix element, and the populations on the sites slowly equilibrate due to the effect of coherence loss, $\rho^{(00)}_t,\rho^{(11)}_t\to\rho^{(ii)}_\infty$,
\begin{equation}
    \difft\rho^{(00)}_t = -i\Omega  \left(\rho^{(01)}_t - \rho^{(10)}_t\right) = 2\Omega\Im\left[\rho^{(01)}_t\right] \,.
\end{equation}

For the example reported in \cref{fig:hakenstrobl_SH_dynamics}, we used the following system parameters: $\varepsilon_0 = 0$, $\varepsilon_1 = 1$, $\Omega = 2$, $\Gamma_0 = \Gamma_1 = 0.5$, and setting $\hbar=1$.

\section{Step-by-step normalization of the linear SSE}\label{app:normalizationSSE}
For the sake of clarity, we show all the steps to write the normalized SSE for a white noise-driven system and, in a similar fashion, the derivation of the associated Lindblad QME.
Starting from the generic white noise SSE in \cref{eq:genericWhiteSSE}, namely
\begin{equation}
        \diff\psi_t = A\psi_t\diff t + \gamma B\psi_t\diff W_t \,,
\end{equation}
the normalization condition is $\mathbb{E}\left[\diff\big(\psi_t\daggah\psi_t\big)\right]=0$.
Then, the differential of the product, 
expanded to 
\begin{equation}
    \diff\big(\psi_t\daggah\psi_t\big) = \diff(\psi_t\daggah)\psi_t + \psi_t\daggah\diff(\psi_t) + \diff(\psi_t\daggah)\diff(\psi_t) \,,
\end{equation}
reads, thanks to the \ito product rule in \cref{eq:ItoProduct}, 
\begin{equation}
\begin{split}
    \diff\big(\psi_t\daggah\psi_t\big)
    = &\; \psi_t\daggah{A\daggah}\psi_t\diff t + \gamma\psi_t\daggah{B\daggah}\psi_t \diff W_t\\
     &+ \psi_t\daggah{A}\psi_t\diff t + \gamma\psi_t\daggah{B}\psi_t\diff{W_t} \\
     &+\gamma^2\psi_t\daggah{B\daggah B}\psi_t\diff{t} \,.
\end{split}
\end{equation}
In the usual braket notation, the argument of the average can be written $\diff{\big(\braket{\psi}{\psi}\big)}$, and the condition can be written equivalently as 
\begin{equation}
\begin{split}
    \diff{\big(\braket{\psi}{\psi}\big)} 
    = &\; \bra{\psi_t }A\daggah\ket{\psi_t}\diff t + \gamma \bra{\psi_t }B\daggah\ket{\psi_t}\diff W_t\\
     &+ \bra{\psi_t}{A}\ket{\psi_t}\diff{t} + \gamma\bra{\psi_t}{B}\ket{\psi_t}\diff{W_t} \\
     &+\gamma^2\bra{\psi_t}{B\daggah B}\ket{\psi_t}\diff{t} \,,
\end{split}
\end{equation}
from which it is easier to see the following recasting of the expression for the differential, and then the normalization condition as 
\begin{equation}
\begin{split}
    \mathbb{E}\left[\diff{\big(\braket{\psi}{\psi}\big)}\right] 
    = &\; \mathbb{E}\left[\bra{\psi_t}{A\daggah+A+\gamma^2 B\daggah B}\ket{\psi_t}\diff{t}\right]  \\
     &+\gamma\mathbb{E}\left[\bra{\psi_t}{B\daggah+B}\ket{\psi_t}\diff{W_t} \right] \,.
\end{split}
\end{equation}
Thanks to the properties of white noise, the second term vanishes, averaging to zero.
To ensure the normalization at all times $t$, we must impose the term in $\diff{t}$ to be zero, and it follows that the total operator acting within the braket must be null,
\begin{equation}
    A\daggah+A+\gamma^2 B\daggah B = 0\,.
\end{equation}
Since the operator $A$ must contain the Hamiltonian of the system, so that the deterministic \schr equation can be recovered in the absence of noise, we define it as
\begin{equation}
    A = -iH+C,
\end{equation}
where the operator $C$ is the correction term needed for the normalization. 
Substituting the definition of $A$ above, the Hamiltonian terms cancel out, and since it is always true that $B\daggah B = (B\daggah B)\daggah$, the renormalization operator reads
\begin{equation}
    C = -\frac{1}{2}\gamma^2 B\daggah B\,,
\end{equation}
leading to the normalized version of the SSE presented in the paper in \cref{eq:Ito_SSE_whitenorm},
\begin{equation}
    \diff\psi_t = \left(-iH -\frac{1}{2}\gamma^2 B\daggah B \right)\psi_t \diff t +\gamma B\psi_t\diff{W_t}\,.
\end{equation}


\bibliography{references}

\end{document}